\documentclass[pre,twocolumn,aps]{revtex4}
\usepackage{amsmath, amssymb}
\usepackage{graphicx}
\usepackage{hyperref}
\usepackage{subcaption}
\usepackage{float}
\usepackage{placeins}
\usepackage{tikz}
\usepackage{xcolor}
\begin{document}

\title{Disorder driven topological phase transitions in 1D mechanical quasicrystals}
\author{Sayan Sircar}
\affiliation{Tata Institute of Fundamental Research, Hyderabad  500046, India}
\date{June 24, 2024}
\begin{abstract}
We examine the transition from trivial to non-trivial phases in a Su-Schrieffer-Heeger model subjected to disorder in a quasi-periodic environment. We analytically determine the phase boundary, and characterize the localization of normal modes using their inverse participation ratio. We compute energy-dependent mobility edges and provide evidence for the emergence of a topological Anderson insulator within specific parameter ranges. Whereas the phase transition boundary is affected by the quasi-periodic modulation, the topologically insulating Anderson phase is stable with respect to the chiral disorder in a quasi-periodic setup. Additionally, our results also uncover a re-entrant topological phase transition from non-trivial to trivial phases for certain values of quasi-periodic modulation with fixed chiral disorder. 
\end{abstract}

\maketitle

\section{INTRODUCTION}
The discovery of topological insulators \cite{RevModPhys.82.3045, RevModPhys.83.1057} in condensed matter physics led to considerable efforts to explore similar phases in several acoustic \cite{PhysRevLett.114.114301} and elastic systems \cite{PhysRevB.96.134307, PhysRevB.98.094302, CHEN201822} with growing applications in photonic systems \cite{Ma_2016, RevModPhys.91.015006}. The bulk-edge correspondence can predict the topological edge states on the boundary of the system, by characterizing the bulk of the material using a topological invariant \cite{Prodan_2016}. Such edge states are protected by internal symmetries and are immune to certain types and levels of disorder \cite{Prodan_2016}. Experimentally, topologically protected conducting edge states in superconducting qubit chains were realized \cite{PhysRevA.98.012331} and proposed in Kitaev materials \cite{timoshuk2023quantum}.
For any clean (with no disorder) system in any spatial dimension, such phase transitions are abrupt. The system suddenly changes from a topological phase hosting conducting edge states to a trivial phase without any conducting edge states. Disorder disrupts the periodicity of the system, which can compromise its topological properties. However, the emergence of Topological Anderson Insulators (TAIs) indicates that disorder can facilitate a transition from topologically trivial to non-trivial states \cite{PhysRevLett.102.136806, PhysRevB.80.165316, PhysRevLett.103.196805, PhysRevB.89.224203, PhysRevLett.113.046802}.
Disorder in real materials typically follows Anderson-type behavior \cite{PhysRev.109.1492,10.1063/1.3206091}, where a random potential arises from a finite concentration of impurities. In three dimensions, a critical potential value $\delta$ \cite{PhysRevLett.42.673} marks the onset of a transition. In one dimension, all states remain localized for small and finite $\delta$, provided the system is large enough.
\par
The emergence of twisted bilayer graphene \cite{Gonçalves_2022} introduces a phenomenon known as quasiperiodic disorder. This has practical applications, at specific incidence angles, electrons in graphene layer  experiences an in-commensurate potential, influenced by the second twisted layer.
The history of quasi-periodicity in electronic systems, goes back to studies by
Aubry and Andre \cite{Domínguez-Castro_2019,aubry1980analyticity}. Aubry and Andre proposed a one-dimensional tight-binding model in which electrons are influenced by a sinusoidal electrostatic potential that is incommensurate with the lattice of the system. They showed that a metal-insulator delocalization-localization transition takes place for $\delta^{AA}=2J$, where $J$ is the nearest-neighbor electron hopping amplitude, where all eigenmodes are either exponentially localized if $\delta^{AA}>2J$, or extended plane waves if $\delta^{AA}<2J$. These results differ from those of Anderson localization in one-dimensional systems. \par
The Su-Schrieffer-Heeger (SSH) model is the best and most transparent model for investigating many such transitions. 
A mechanical version of SSH model is shown in Fig.~\ref{fig:qw}.
The topological features of the model has two topological phases characterized by winding numbers $\nu=0,1$, indicating trivial and non-trivial phases. Recent studies have shown that both chiral-preserving \cite{PhysRevResearch.3.033012} and chiral-breaking \cite{PhysRevB.107.035113} disorder can induce topological phase transitions in SSH-based systems.
Introducing long-range hopping across the chain make it possible to host more than one conducting state per edge \cite{PhysRevB.109.035114}.
Topological phases arise from various internal symmetries of the model \cite{Ryu_2010}, such as chiral, time-reversal, and particle-hole symmetries, which also apply to a non-Hermitian variant featuring more complex symmetry like anti-PT symmetry \cite {PhysRevB.103.235110}. 
When exposed to quasiperiodicity, such a model exhibits non-trivial localization, which is absent in the Aubry-André (A-A) description. This includes the presence of energy-dependent mobility edges and critical phases across various disorder strengths \cite{Tang_2023, PhysRevLett.126.106803,PhysRevLett.131.176401}.
The equation governing the $A-A$ model describing the application of sinusoidal varying on-site potential to each lattice site in a periodic lattice system is:
\begin{equation}
    t(\psi_{j+1}+\psi_{j-1})+\Delta\cos(2\pi \beta j +\phi)\psi_{j}=E\psi_{j}.
\end{equation}
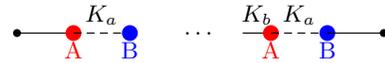
\begin{figure}[ht!]
\centering
\begin{tikzpicture}[scale=0.75]

\coordinate (F1) at (0,0);
\coordinate (A1) at (1,0);
\coordinate (B1) at (2,0);

\coordinate (A_end) at (4.5,0);
\coordinate (B_end) at (5.5,0);
\coordinate (F2) at (6.5,0);

\fill (F1) circle (2pt);
\fill (F2) circle (2pt);

\draw (F1) -- (A1);
\fill[red] (A1) circle (4pt) node[below] {A};
\draw[densely dashed] (A1) -- (B1) node[midway, above] {$K_{a}$};
\fill[blue] (B1) circle (4pt) node[below] {B};

\node at (3.25,0) {\ldots};

\coordinate (A_ellipsis) at (4,0);
\draw (A_ellipsis) -- (A_end) node[midway, above] {$K_{b}$};
\fill[red] (A_end) circle (4pt) node[below] {A};
\draw[densely dashed] (A_end) -- (B_end) node[midway, above] {$K_{a}$};
\fill[blue] (B_end) circle (4pt) node[below] {B};
\draw (B_end) -- (F2);

\end{tikzpicture}
\caption{Schematic depiction of a mechanical SSH chain, with dashed lines indicating intra-cellular springs ($K_{a}$) and solid lines indicating inter-cellular springs ($K_{b}$). A single unit cell is from masses $A$ and $B$. The black solid circles at the ends indicate fixed ends.}
\label{fig:qw}
\end{figure}
where $\Delta$ is the strength of quasi-periodic modulation, $t$ is the nearest neighbor hopping amplitude, $\phi$ is the phase parameter, which plays a crucial role in commensurate modulation, $\beta$ is an irrational number which plays the same role as magnetic field strength in Hofstafder butterfly setting. In this model, quasi-periodicity can induce abrupt metal-insulator phase transition at $\Delta=2t$ \cite{Tang_2023}, and also the model is self-dual at this point. \par
This problem aims to extend concepts from hopping Hamiltonians in quantum models to a simpler classical framework, incorporating disorder. The goal is to demonstrate that topological phase transitions are not exclusive to quantum systems but can also be effectively realized in classical systems. Exploring these phases classically will facilitate a clearer study of disordered topological insulators. \par
The paper is structured as follows. In Sec.~\ref{sec:2}, we introduce the mechanical SSH model, discuss various notations, and explore the topological properties of the clean system (without disorder or unit cell-dependent modulation). In Sec.~\ref{sec:3}, we examine the impact of $AA$ modulation and discrete disorder on the topological properties and discuss the its impact on localization properties of the eigenstates. In Sec.~\ref{sec:4}, we identify and explain the presence of mobility edges, culminating in the recognition of the TAI phase in a disordered model with a quasi-periodic background.

\section{MECHANICAL MODEL FOR SSH CHAIN}\label{sec:2}
This section introduces our SSH mechanical model, defines the notation, and examines the topological properties of the model from a real-space perspective.

\subsection{Model and dynamical matrix}
We consider a 1D finite chain with fixed boundaries, composed of N unit cells, each containing two equal masses. Each mass is designated by a sub-lattice index $\alpha={A,B}$ and a unit cell index $j\in[1,N]$. The system features two types of spring constants: $K_{a}$ for intra-cellular and $K_{b}$ for inter-cellular interactions, analogous to the hopping parameters in the SSH model, as illustrated in Fig.~\ref{fig:qw}. By adjusting these parameters, various topological phases can be realized, similar to the quantum SSH model. The equations of motion for each mass can be derived from Newton's second law or through the Euler-Lagrange equation, commonly used in normal modes analysis. These equations can be expressed compactly using matrix notation. We also assume that each mass has a harmonic displacement in time as 
\begin{equation}
    u_{j}(t)=u_{j}e^{-i\omega_{j}t},
\end{equation}
because the normal modes have a well defined frequency. We can transform the second-order equations of motion into a linear matrix equation. This matrix, known as the dynamical matrix, has eigenvectors and eigenvalues that represent the normal modes of the system and their corresponding frequencies,
\begin{equation}
    DU=\omega^{2}U (m=1),
\end{equation}
Where $D$ is the dynamical matrix of dimension $2N\times2N$, and $U$ is the displacement vector associated with the displacement of each mass $U=[u_{1}^{A},u_{1}^{B},u_{2}^{A},u_{2}^{B},\cdot \cdot \cdot,u_{N}^{A},u_{N}^{B}]^{T}$ with $u_{j}^{\alpha}$ is the displacement of mass with unit cell index $j$ and sub-lattice index $\alpha=A/B$. The explicit form of the matrix is 
\begin{equation}\label{eq:dynamicalmat1}
  D= \begin{bmatrix}
K_{a}+K_{b} & -K_{a} & \dots & 0 \\
-K_{a} & K_{a}+K_{b} & -K_{b} & \dots \\
\vdots & \vdots & \ddots & -K_{a} \\
\dots & \dots & -K_{a} & K_{a}+K_{b} \\
\end{bmatrix}. 
\end{equation}

The matrix has a tri-diagonal structure. Unlike the original SSH Hamiltonian, where changes in inter-cellular hopping only affect the off-diagonal elements, modifications in inter-cellular stiffness also impact the diagonal elements in our mechanical scenario. Since all diagonal elements are equal,it retains the chirality of the dynamical matrix. These non-zero diagonal terms shift the eigenvalue spectrum to $\omega_{0}^{2}=K_{a}+K_{b}$, with all other eigenvalues measured relative to this value. We can define the matrix $D^{'}=D-\omega_{0}^{2}I$, which satisfies an anti-commutation relation with the chiral operator $\Gamma$, centering the eigenvalue spectrum around $\omega_{0}^{2}=0$,
\begin{equation}
    \Gamma(D-\omega_{0}^{2}I)+(D-\omega_{0}^{2}I)\Gamma=0,
\end{equation}
where $I$ denotes the $2N\times2N$ identity matrix, and $\Gamma$ represents the chiral operator. The matrix depiction of the chiral operator for the spring mass chain with fixed boundary conditions is: 
{\small
\begin{equation}\label{eq:chiralmatrix}
  \Gamma=  \begin{bmatrix}
1 & 0 & \dots & 0 \\
0 & -1 & 0 & \dots \\
\vdots & \vdots & \ddots & 0 \\
\dots & \dots & 0 & 1 \\
\end{bmatrix} .  
\end{equation}
}
\begin{figure*}[htbp]
    \centering
    \includegraphics[keepaspectratio, width=0.45\textwidth]{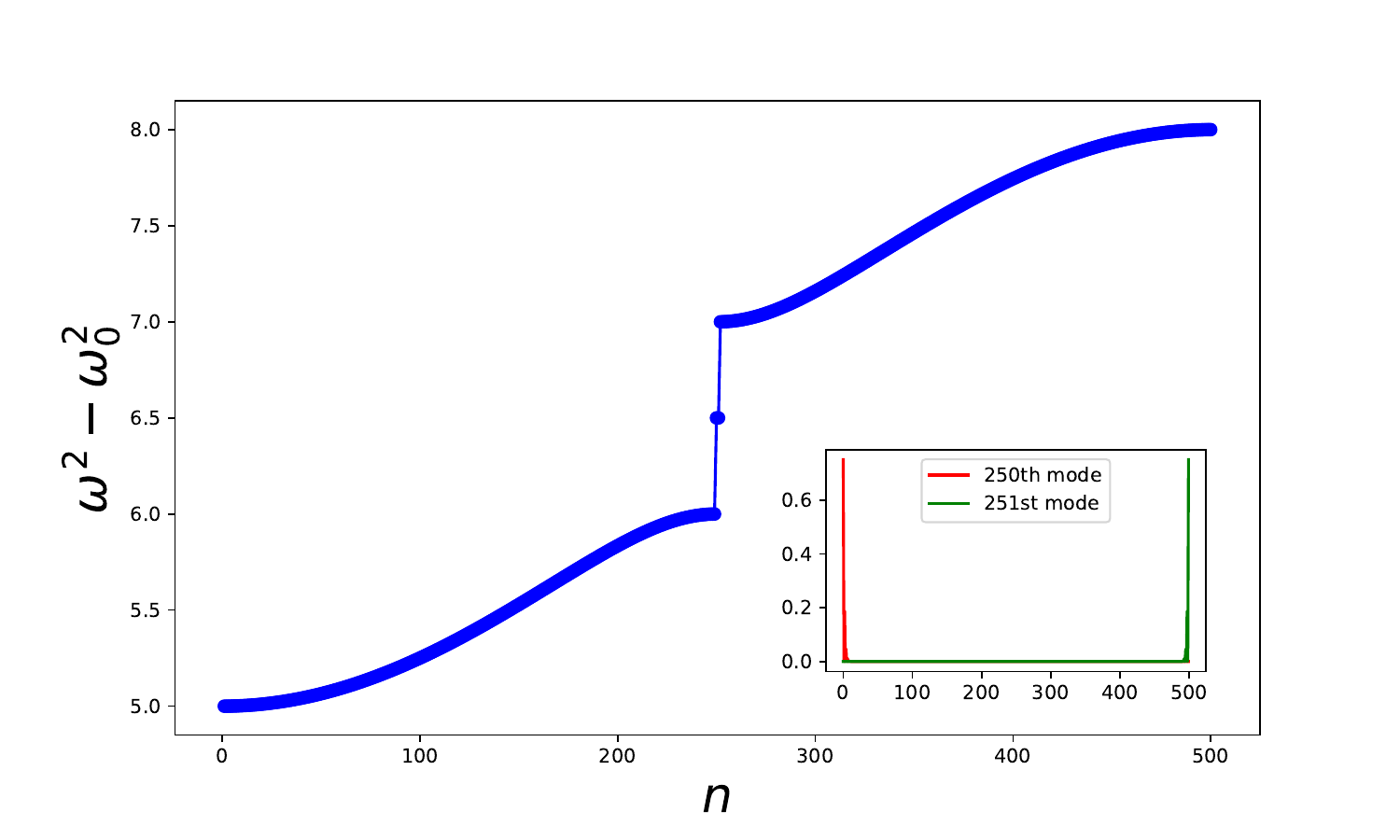}
    \includegraphics[keepaspectratio, width=0.45\textwidth]{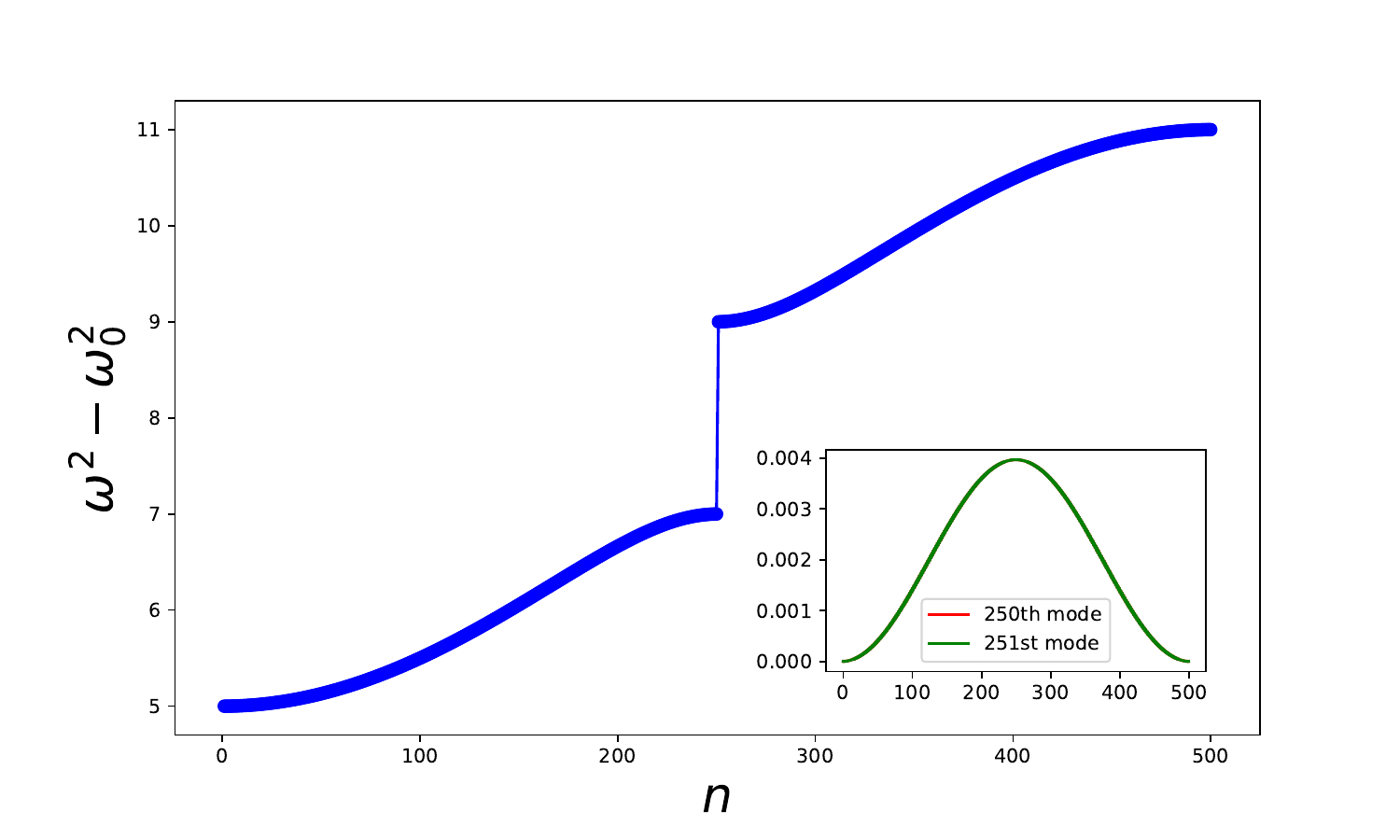}
     \put(-402,100){\textbf{(a) }}
     \put(-170,100){\textbf{(b)}}
     \\
     \caption{Energy versus mode number plot for a 1D spring-mass chain in the case of a clean system. (a) System indicating mid-frequency states. Eigenvalues are computed from the dynamical matrix provided in Eq.~(\ref{eq:dynamicalmat1}). The probability distribution of the mid-gap states is displayed as an inset, indicating localized modes at the boundaries. The color indicates the chirality of the edge modes. (b) System without mid-frequency states. The inset reveals that the probability distribution of mid-gap states extends across the entirety of the bulk, devoid of edge modes, with the loss of chirality, signifying a topologically trivial phase.}
    \label{fig:combined_vector_plot2}
\end{figure*}
\subsection{Characterizing the topology in real space}
For the topological analysis of quantum tight-binding hopping models, winding number calculations in momentum space \cite{CHEN201822} has been performed. In large system sizes, translational symmetry holds within the bulk but is disrupted near the boundaries. In the absence of disorder, our system retains translational symmetry in the bulk; however, disorder introduces disruptions to this symmetry. The analysis can be conducted in real space \cite{PhysRevResearch.3.033012,PhysRevLett.113.046802}, utilizing the concept of a local topological marker (LTM) to characterize the topology \cite{meier2018observation}. The LTM assigns a local value to each unit cell, and when averaged throughout the interior of the system, it converges to the winding number of the periodic system. A key benefit of the LTM is its ability to handle disordered systems as well \cite{PhysRevLett.113.046802,PhysRevResearch.3.033012}.

For computing LTMs for a chain of $N$ unit cells, we have constructed two $N \times N$ matrices of eigenvectors from the dynamical matrix. The matrix $U_{-}=[U_{1},U_{2},U_{3},\dots,U_{N}]$ contains eigenvectors below the zero-energy band gap, while $U_{+}=[U_{N+1},U_{N+2},\dots,U_{2N}]$ holds those above it. Using $U_{+}$ and $U_{-}$, we can define projectors for the bands as $P_{-}=U_{-}U_{-}^{T}$ and $P_{+}=U_{+}U_{+}^{T}$, leading to a flat-band Hamiltonian \cite{PhysRevResearch.3.033012}. This Hamiltonian, topologically equivalent to $D$, is expressed as $Q=P_{+}-P_{-}$ and is decomposed into $Q_{AB}+Q_{BA}=\Gamma_{A}Q\Gamma_{B}+\Gamma_{B}Q\Gamma_{A}$ for the $A$ and $B$ sub-lattices,
\begin{equation}
    \Gamma_{A/B}= \begin{bmatrix}
1/0 & 0 & 0 & 0 & \dots\\
0 & 0/1 & 0 & 0 & \dots \\
0 & 0 & 1/0 & 0 & \dots\\
\dots & \dots & \dots & \dots & \dots \\
\end{bmatrix} ,
\end{equation}
are the sub-lattice projectors and $\Gamma=\Gamma_{A}-\Gamma_{B}$ is the chiral operator. Finally the definition of LTM \cite{meier2018observation,PhysRevResearch.3.033012} is 
\begin{equation}\label{eq:LTM1}
    \nu(k)=\frac{1}{2}\sum_{\alpha=A,B}{(Q_{BA}[X,Q_{AB}])_{k\alpha,k\alpha}+(Q_{AB}[Q_{BA},X])_{k\alpha,k\alpha}},
\end{equation}

where the position operator $X$, with dimensions dim($X$)=$2N\times 2N$, maps masses to their unit cell positions from the center of the system: $X=[-N,-N,-(N-1),-(N-1),\dots,(N-1),(N-1)]$ on the diagonal. The LTM converges to the winding number, which takes on two values: $\nu=0$ for the topologically trivial phase and $\nu=1$ for the non-trivial phase. We realise each by tuning the ratio of $K_{a}$ to $K_{b}$, as illustrated in Fig.~\ref{fig:combined_vector_plot2}. The ratio $\frac{K_{b}}{K_{a}}>1$ indicates a non-trivial phase, while $\frac{K_{b}}{K_{a}}<1$ indicates a trivial phase. In the topological phase, the system exhibits conducting edge states, which are absent in the trivial phase or with periodic boundary conditions (PBC).
\section{Dynamical matrix of the spring mass chain with AUbry-Andr'e quasi-periodic modulation in intercellular spring constants and discrete disorder in intracellular spring constants}\label{sec:3}

\subsection{Equations of motion (Euler-Lagrange equations)}
We investigate a nearest-neighbour mechanical model affected by $AA$ modulation and discrete disorder. The modulation, defined by a sinusoidal term with an incommensurate wavelength—arising from the choice of the irrational number $\beta$, which is the inverse of the golden ratio $g^{-1}=\frac{2}{1+\sqrt{5}}$—is central to our study. Our goal is to understand how induced discrete disorder in intra-cellular spring stiffness and quasi-disorder from $AA$ modulation influence the topological properties of the system, employing the discussed local topological marker formalism.
The equations of motion of our model for type $A$ mass in $j$th unit cell will be 
\begin{equation}
    \begin{split}
   \ddot{u}^{A}_{j} &= -[K_{a} + \delta_{j}^{d} + K_{b} + \delta^{AA} \cos(2\pi \beta j + \phi)] u^{A}_{j} \\
   &\quad + [K_{a} + \delta_{j}^{d}] u^{B}_{j} + [K_{b} + \delta^{AA} \cos(2\pi \beta j + \phi)] u^{B}_{j-1}.
\end{split}
\end{equation}
Similarly for type $B$ mass in $j$th unit cell,
\begin{equation}
    \begin{split}
   \ddot{u}^{B}_{j} &= -[K_{a} + \delta^{d}_{j} + K_{b} + \delta^{AA} \cos(2\pi \beta (j+1) + \phi)] u^{B}_{j} \\
   &\quad + [K_{a} + \delta^{d}_{j}] u^{A}_{j} + [K_{b} + \delta^{AA} \cos(2\pi \beta (j+1) + \phi)] u^{A}_{j+1},
\end{split}
\end{equation}
where $\delta^{AA}$ is $AA$ amplitude, and $\delta^{d}_{j}$ (with $j\in[1,N]$ as unit cell index) is a discrete disorder term defined as 
\begin{equation}
k_{j}=k_{j}+\delta_{j}^{d}=K_{a}+\delta_{d}\epsilon_{j}, 
\end{equation} for $j$ even
where $\delta_{d}$ represents the disorder strength regarding intra-cellular spring stiffness, consistent with the distribution of $\epsilon$ (independent and identically distributed random variables drawn from the range $[-1,1]$). The perturbation on the spring stiffness leads to terms on the diagonal, unlike the original SSH model. This off-diagonal perturbation alters both the diagonal and off-diagonal terms, disrupting the chirality of the dynamical matrix. To preserve the symmetry (chiral), local springs are included for each mass.

\begin{figure}[t!]
    \centering
    \includegraphics[width=\linewidth]{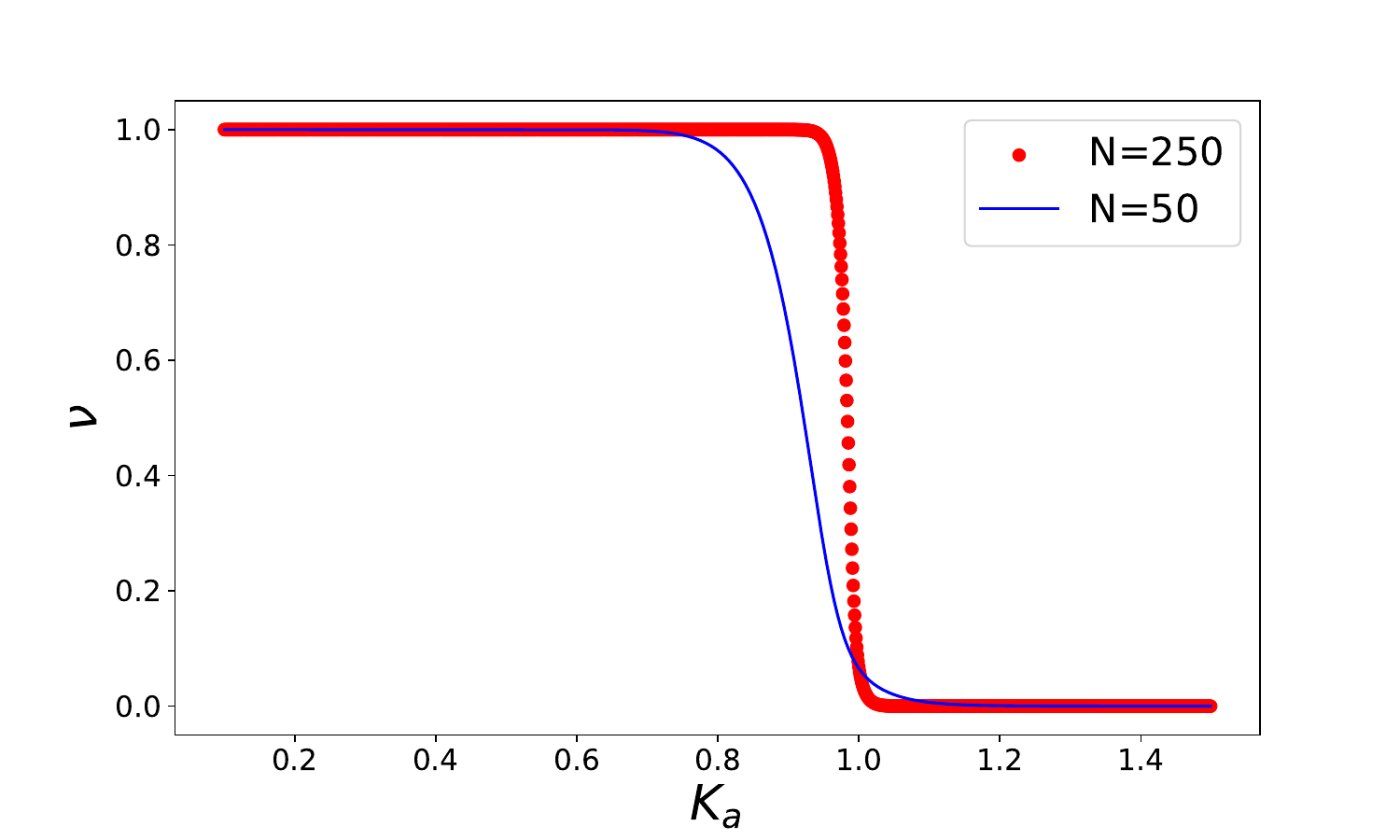}
    \caption{Evolution of LTM plot with intra-cellular spring stiffness for a single unit cell away from the boundary without disorder or modulation. The graph shows plots for various system sizes. It is noted that there is a distortion in the sudden change of winding number during phase transition due to finite size effect. In a system of infinite size with translational invariance, LTM converges to winding number. The plot demonstrates that as system size grows, there is a shift in LTM development from gradual to sudden, in line with theoretical predictions.}
    \label{fig:LTMEVOLUTIONCLEAN}
\end{figure}
Including the on-site spring at each mass, in order to preserve the chiral symmetry,
the equations of motion of our model for type $A$ mass in $j$th unit cell will be 
\begin{equation}
    \begin{split}
   \ddot{u}^{A}_{j} &= -[K_{a} + \delta^{d}_{j} + K_{b} + \delta^{AA} \cos(2\pi \beta j + \phi)+K^{0,A}_{j}] u^{A}_{j} \\
   &\quad + [K_{a} + \delta^{d}_{j}] u^{B}_{j} + [K_{b} + \delta^{AA} \cos(2\pi \beta j + \phi)] u^{B}_{j-1}.
\end{split}
\end{equation}
Similarly the equation of motion for type$B$ mass in $j$th unit cell will be 
\begin{equation}
    \begin{split}
   \ddot{u}^{B}_{j} &= -[K_{a} + \delta^{d}_{j} + K_{b} + \delta^{AA} \cos(2\pi \beta (j+1) + \phi)+K^{0,B}_{j}] u^{B}_{j} \\
   &\quad + [K_{a} + \delta^{d}_{j}] u^{A}_{j} + [K_{b} + \delta^{AA} \cos(2\pi \beta (j+1) + \phi)] u^{A}_{j+1},
\end{split}
\end{equation}
where $K^{0,A}_{j}=K_{0}+\delta^{AA}[1-\cos(2\pi \beta j+\phi)]-\delta^{d}_{2j}$ and $K^{0,B}_{j}=K_{0}+\delta^{AA}[1-\cos(2\pi \beta (j+1)+\phi)]-\delta^{d}_{2j}$.
The choice of on-site spring stiffness preserves the chirality of the matrix and ensures all eigenvalues of the problem are positive. Assuming a normal mode solution, the two equations above can be expressed as:
\begin{equation}
\begin{aligned}
\omega^{2}u^{A}_{j} &= (K_{a} + K_{b} + \delta^{AA} + K_{0})u^{A}_{j} \\
                     &\quad -(K_{a} + \delta^{d}_{j})u^{B}_{j} \\
                     &\quad -(K_{b} + \delta^{AA}\cos(2\pi \beta j + \phi))u^{B}_{j-1},
\end{aligned}
\end{equation}
\vspace{0.25cm}   similarly for type $B$ mass one has 
\begin{equation}
\begin{aligned}
\omega^{2}u^{B}_{j} &= (K_{a} + K_{b} + \delta^{AA} + K_{0})u^{B}_{j} \\
                     &\quad -(K_{a} + \delta^{d}_{j})u^{A}_{j} \\
                     &\quad -(K_{b} + \delta^{AA}\cos(2\pi \beta (j+1) + \phi))u^{A}_{j+1}.
\end{aligned}
\end{equation} 
The above two equations can be expressed succinctly using a dynamical matrix similar to Eq.~\eqref{eq:dynamicalmat1},

\FloatBarrier




\begin{figure*}[htbp]
    \centering
    \includegraphics[keepaspectratio, width=0.45\textwidth]{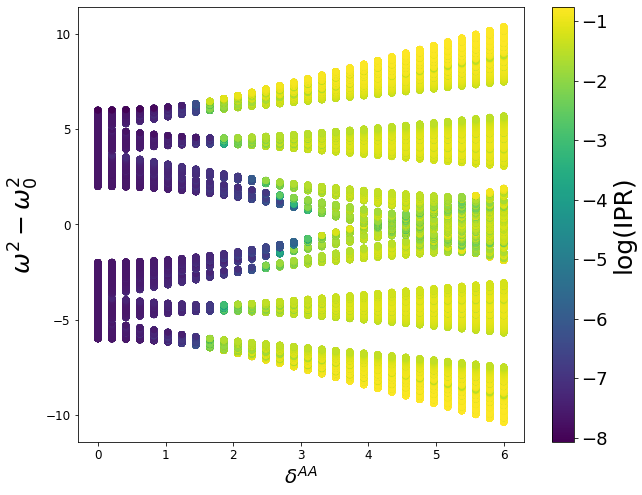}
    \includegraphics[keepaspectratio, width=0.45\textwidth]{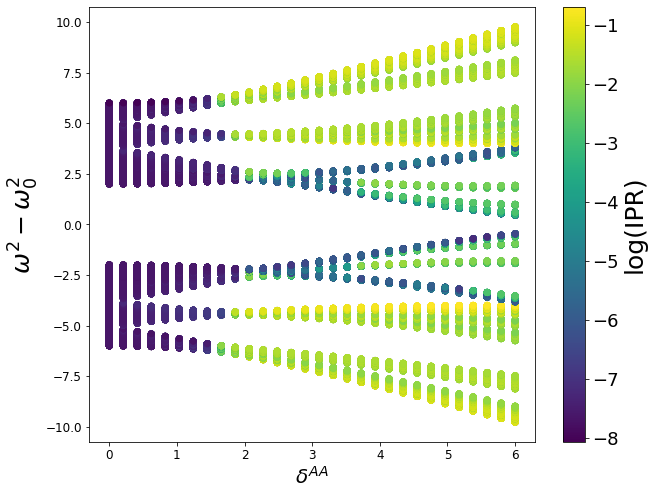}
     \put(-402,160){\textbf{(a) }}
     \put(-170,157){\textbf{(b)}}
     \\
    \includegraphics[keepaspectratio, width=0.45\textwidth]{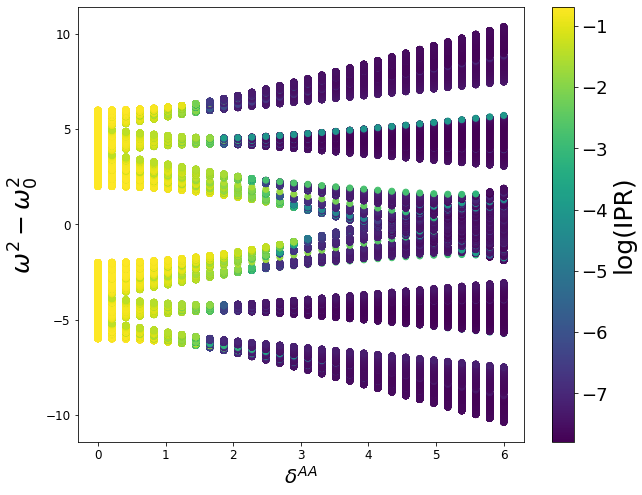}
    \includegraphics[keepaspectratio, width=0.45\textwidth]{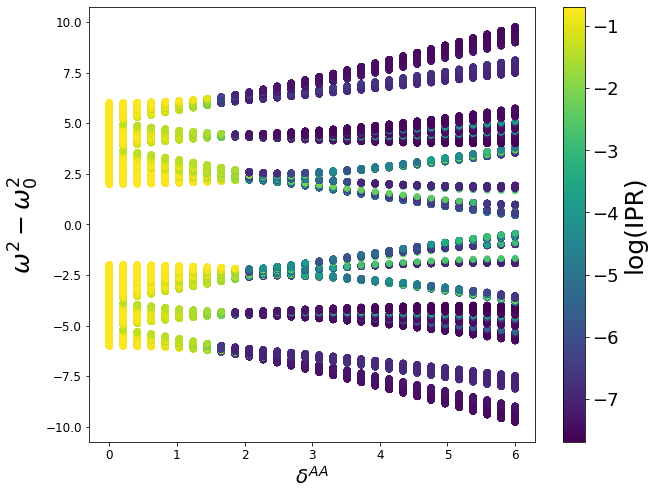}
    \put(-400,160){\textbf{(c) }}
    \put(-170,157){\textbf{(d) }}
    \\
    \caption{$\ln(IPR)$ as a function of the eigenvalue spectrum $\omega^{2}-\omega_{o}^{2}$ for $N=1597$ unit cells, $\delta_{d}=0$, $\phi=0$. For $K_{a}<K_{b}$, the transition from extended phase to localized phase in real space and localized phase to extended phase in momentum space is much more sharp, while for $K_{a}>K_{b}$, there are certain regions where the transition is highligted (i.e. either converges to $0$ or $1$), and regions where the IPR of the eigenstate does not converge to $1$ or $0$. This provides an indication of the presence of a critical phase, which we have quantitatively verified in Fig.~\ref{fig:combined_vector_plot3}, Fig.~\ref{fig:combined_vector_plot4}. (a): $K_{a}<K_{b}$ in real space, (b): $K_{a}>K_{b}$ in real space, (c): $K_{a}<K_{b}$ in momentum space, (d): $K_{a}>K_{b}$ in momentum space.}
    \label{fig:combined_vector_plot2}
\end{figure*}

\begin{widetext}
{\small
\begin{equation}\label{eq:dynamicalmat2}
  D=\begin{bmatrix}
K_{a}+K_{b}+\delta^{AA}+K_{0} & -K_{a}-\delta^{d}_{2} & \dots & -[K_{b}+\delta^{AA}\cos(2\pi\beta+\phi)] \\
-K_{a}-\delta^{d}_{2} & K_{a}+K_{b}+\delta^{AA}+K_{0} & -[K_{b}+\delta^{AA}\cos(4\pi\beta+\phi)] & \dots \\
\vdots & \vdots & \ddots & -K_{a}-\delta^{d}_{N} \\
-[K_{b}+\delta^{AA}\cos(2\pi\beta+\phi)] & \dots & -K_{a}-\delta^{d}_{N} & K_{a}+K_{b}+\delta^{AA}+K_{0} \\
\end{bmatrix}   .
\end{equation}
}
\end{widetext}

The matrix has dimensions $2N\times 2N$.
For true quasi-periodicity, working with an infinitely sized system is required. To mimic this, one can apply periodic boundary conditions on the lattice and define an appropriate rational approximation $\tilde{\beta}=\frac{F_{n-1}}{F_{n}}$, where the numerator and denominator are consecutive Fibonacci sequence terms.

\subsection{ Incommensurate modulation}\label{sec:b}



\begin{figure*}[htbp]
    \centering
    \includegraphics[keepaspectratio, width=0.45\textwidth]{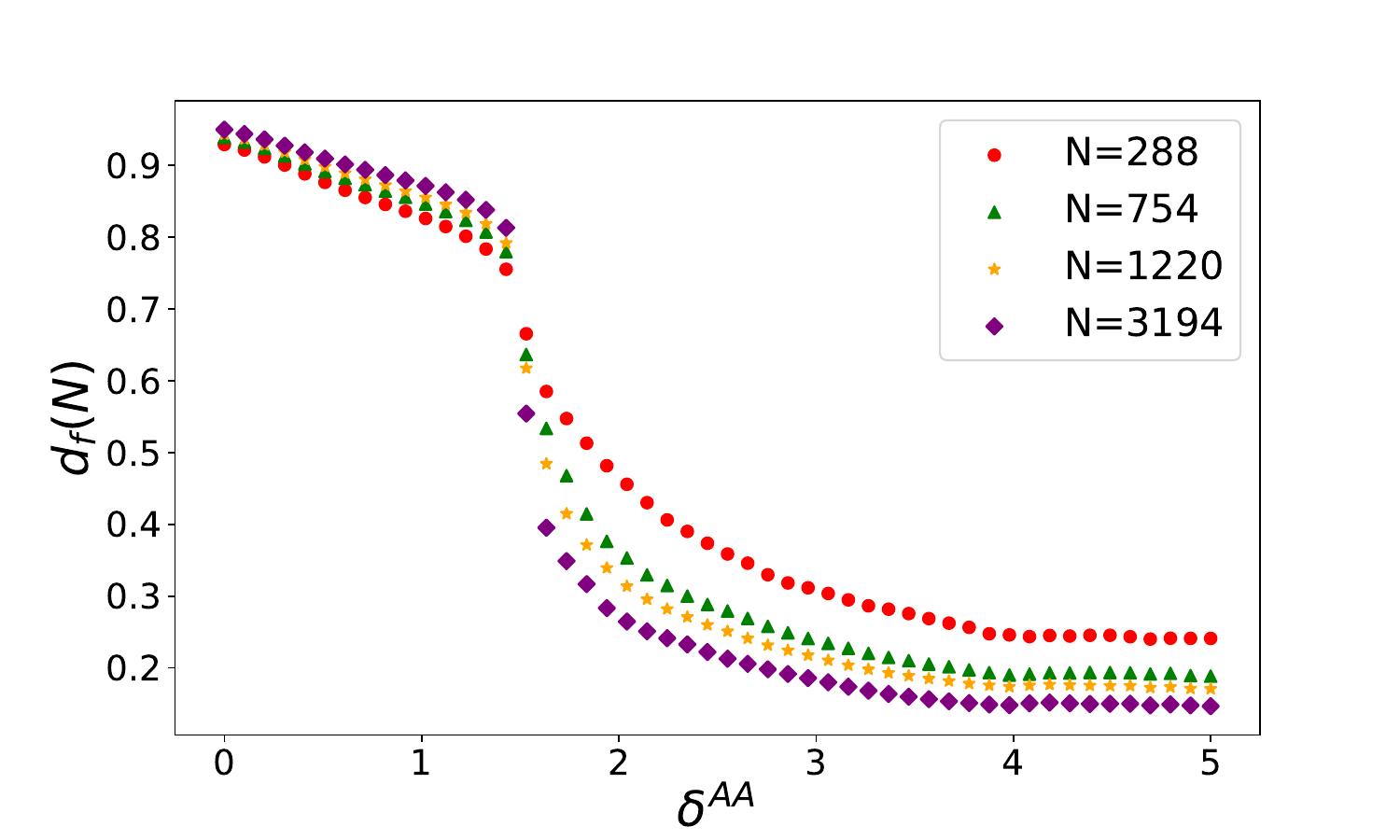}
    \includegraphics[keepaspectratio, width=0.45\textwidth]{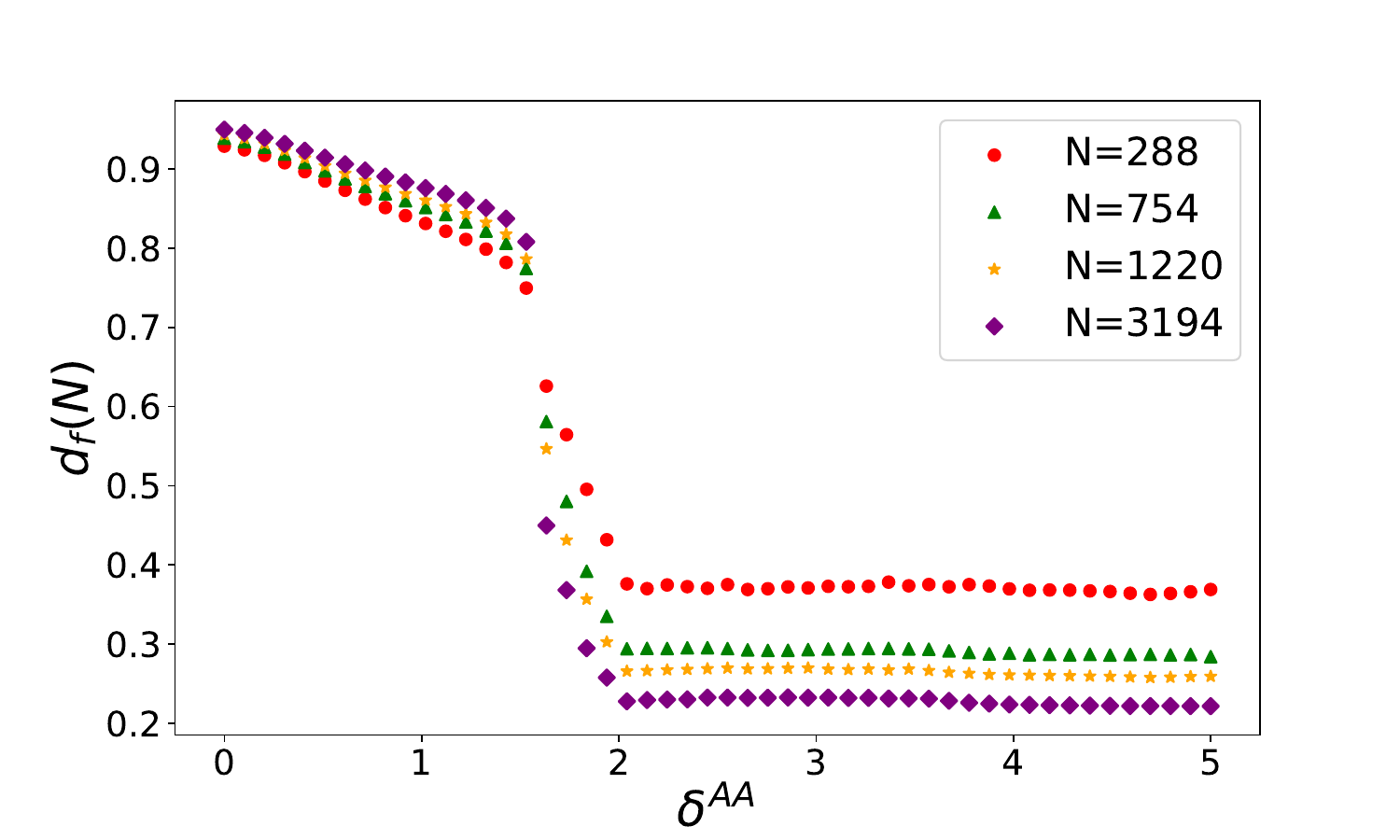}
     \put(-360,110){\textbf{(a) }}
     \put(-145,110){\textbf{(b)}}
     \\
    \includegraphics[keepaspectratio, width=0.45\textwidth]{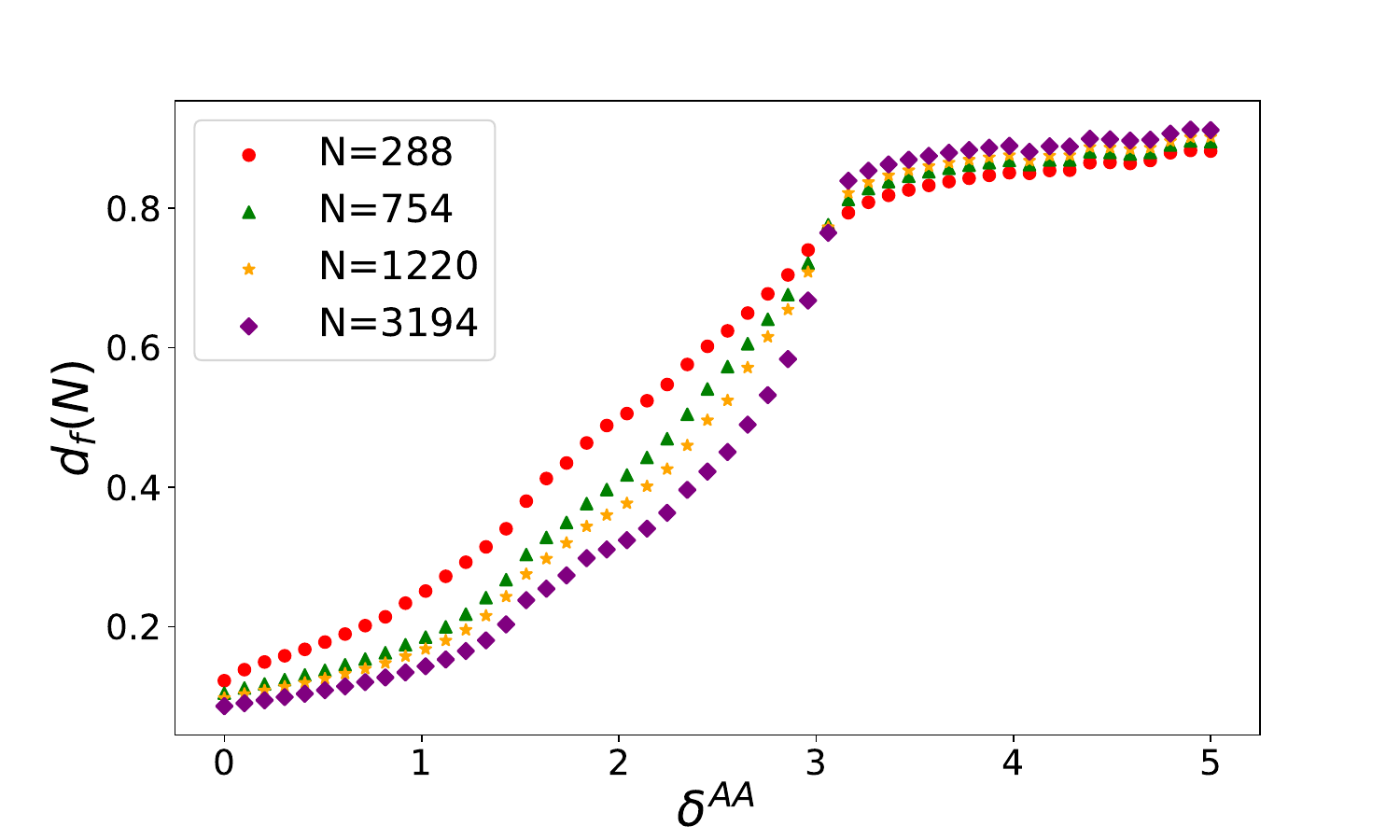}
    \includegraphics[keepaspectratio, width=0.45\textwidth]{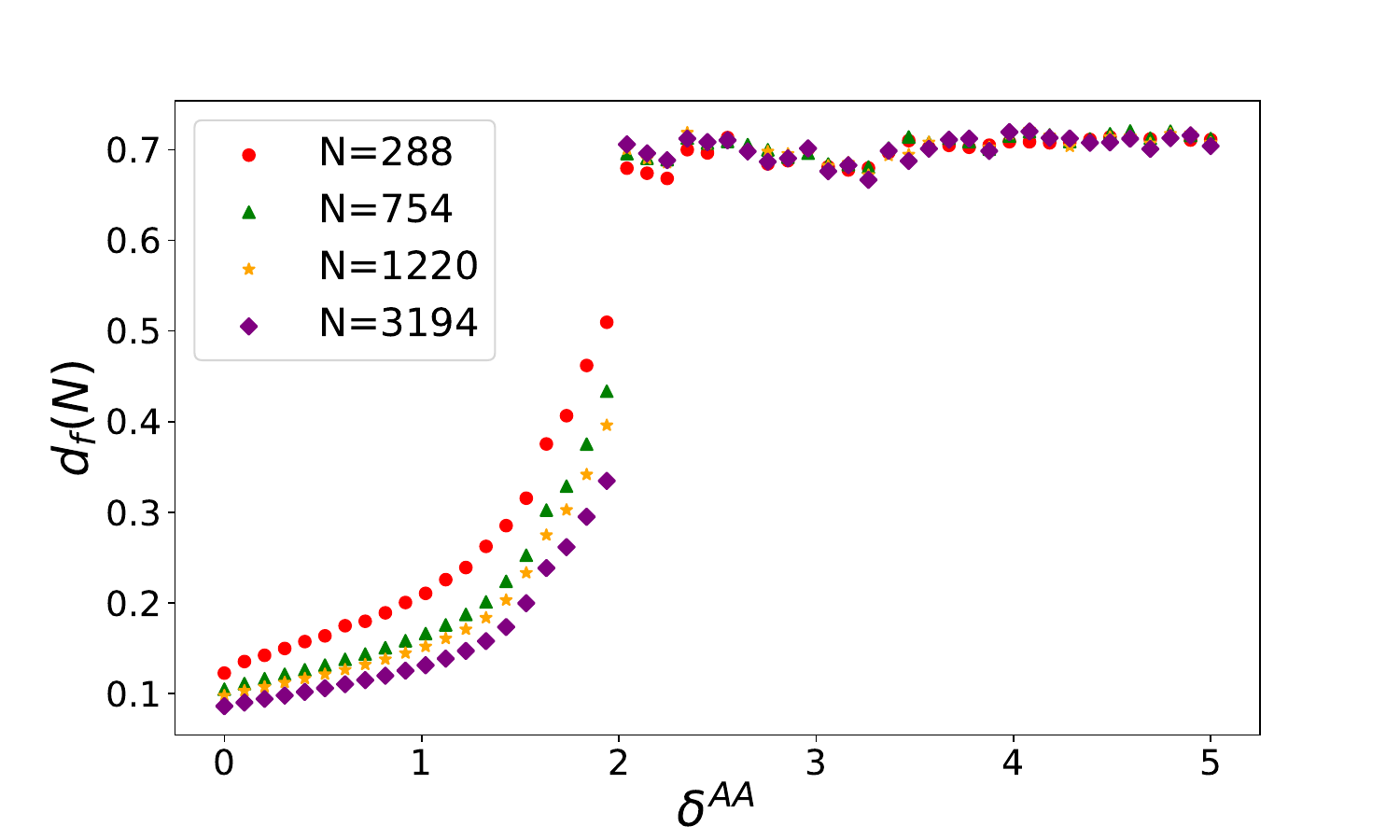}
    \put(-360,110){\textbf{(c) }}
    \put(-145,110){\textbf{(d) }}
    \\
    \caption{Plot of fractal dimension as a function of $\delta^{AA}$ for $\delta_{d}=0$, $\phi=0$ for various system sizes. (a): $K_{b}>K_{a}$ in real space, (b): $K_{a}>K_{b}$ in real space, (c): $K_{b}>K_{a}$ in momentum space, (d): $K_{a}>K_{b}$ in momentum space. In cases (a) and (b) the system begins from an extended phase and transitions quickly into a localized phase as $\delta^{AA}$ increases. However, in case (b), after $\delta^{AA}>2$, the value remains constant with increasing $\delta^{AA}$ dependent of $N$. In cases (c) and (d), the system begins from being in a localized phase and make transitions slowly to an extended phase, but in case (d), for $\delta^{AA}>2$, the value of $d_{f}(N)$ remains constant with increasing $\delta^{AA}$ independent of system size $N$. The nature of the curve, especially in case (d), implies that there is a presence of critical phase in the system with these parameter values.}
    \label{fig:combined_vector_plot3}
\end{figure*}

The topology will be characterised by computing LTM as labelled in Eq.~\eqref{eq:LTM1}, for each $K_{a}$ and $\delta^{AA}$ or vice-versa, by fixing the values of $K_{b}$ and $\delta_{d}$, which converges to the winding number. We have  fixed $K_{b}=2$,$\phi=0$, rational approximation $\tilde{\beta}=\frac{987}{1597}$, $N=1597$ unit cells, $K_{a}\in[0,5]$, and $\delta^{AA}\in[0,6]$. As expected, the topological phase transition occurs when $K_{a}=K_{b}$ in the original SSH model with $\delta^{AA}=0$ \cite{Asb_th_2016}.
Secondly, the behavior of the localization length has also been studied. It is a known fact that in electronic band systems, the topological phase transitions are accompanied by the localization length that diverges at the Fermi level \cite{PhysRevLett.113.046802, LIU2022128004, PhysRevLett.112.206602}. It coincides with the band gap closing. To confirm this in our mechanical model, the localization length $\Lambda$ is calculated analytically and has been seen to diverge in the transition region between the topological phases, where the gap closes. Here the entire eigenvalue spectrum has been measured relative to the middle frequency $\omega_{0}^{2}=K_{a}+K_{b}+K_{0}+\delta^{AA}$, such that the eigenvalue spectrum can have mid-gap near zero energy, when there is a closing of the gap i.e. $\omega^{2}-\omega_{0}^{2}=0$. Hence the equation of motions for states with zero energy, where the gap closes are 
\begin{equation}
    (K_{a} + \delta^{d}_{j})u^{B}_{j}+(K_{b} + \delta^{AA}\cos(2\pi \beta j + \phi))u^{B}_{j-1}=0,
\end{equation}
\begin{equation}
    (K_{a} + \delta^{d}_{j})u^{A}_{j}+(K_{b} + \delta^{AA}\cos(2\pi \beta (j+1) + \phi))u^{A}_{j+1}=0.
\end{equation}
Simplifying the equation for $A$ sub-lattice,
\begin{equation}
    u^{A}_{j+1}=\frac{-(K_{a}+\delta^{d}_{j})}{K_{b}+\delta^{AA}\cos(2\pi \beta (j+1)+\phi)}u^{A}_{j}.
\end{equation}
The above equation leads to recursive relation as shown in Eq.~\eqref{eq:19}



\begin{figure*}[htbp]
    \centering
    \includegraphics[keepaspectratio, width=0.45\textwidth]{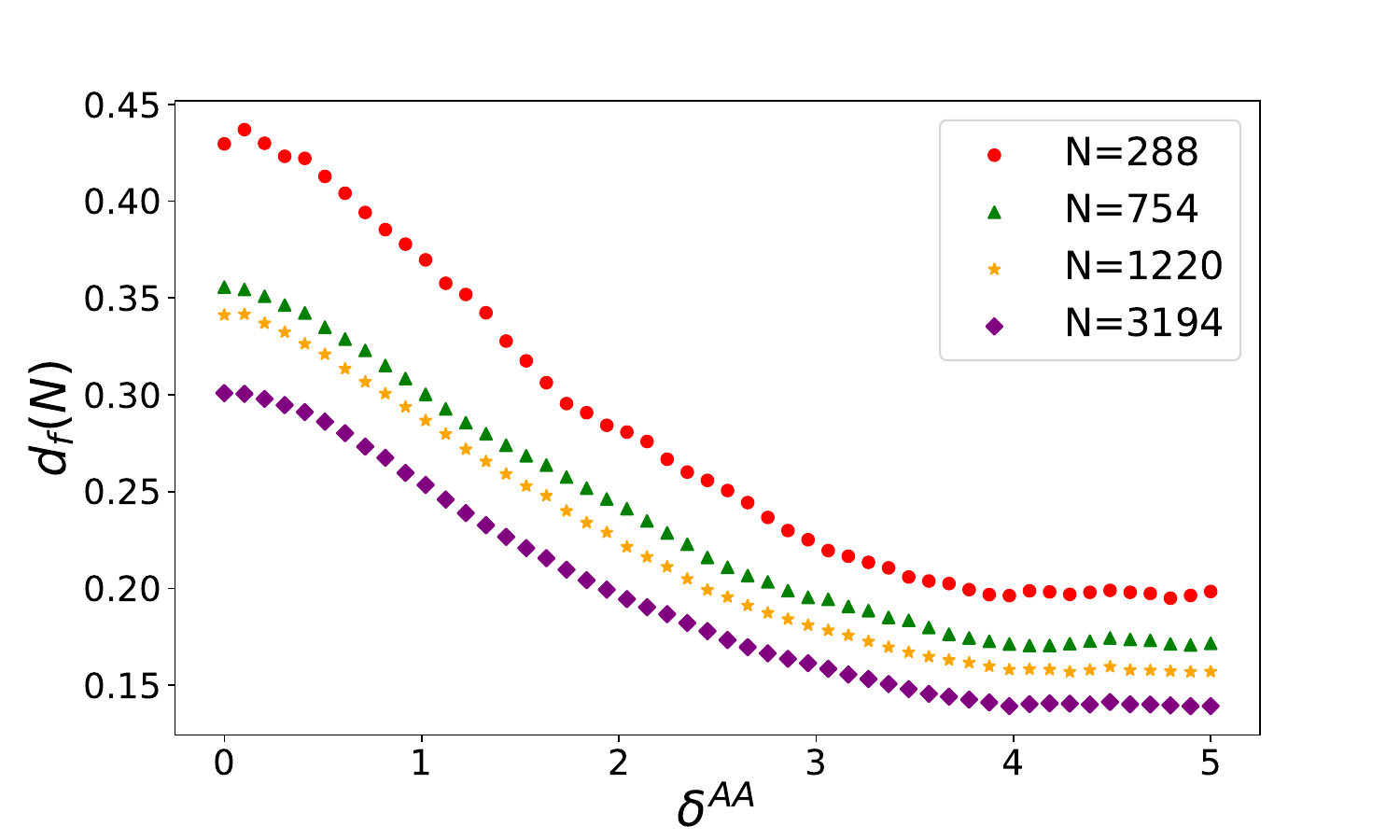}
    \includegraphics[keepaspectratio, width=0.45\textwidth]{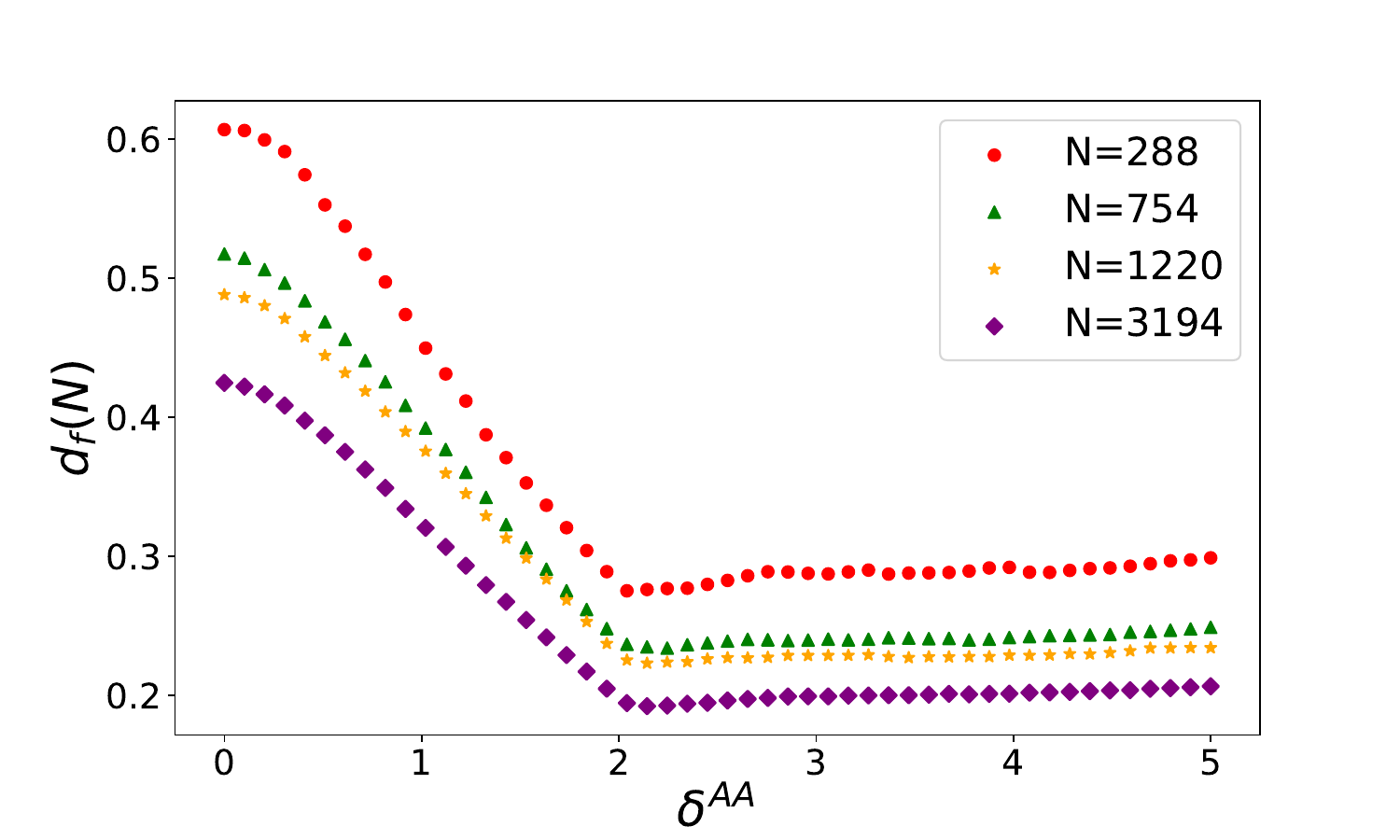}
     \put(-402,110){\textbf{(a) }}
     \put(-160,110){\textbf{(b)}}
     \\
    \includegraphics[keepaspectratio, width=0.45\textwidth]{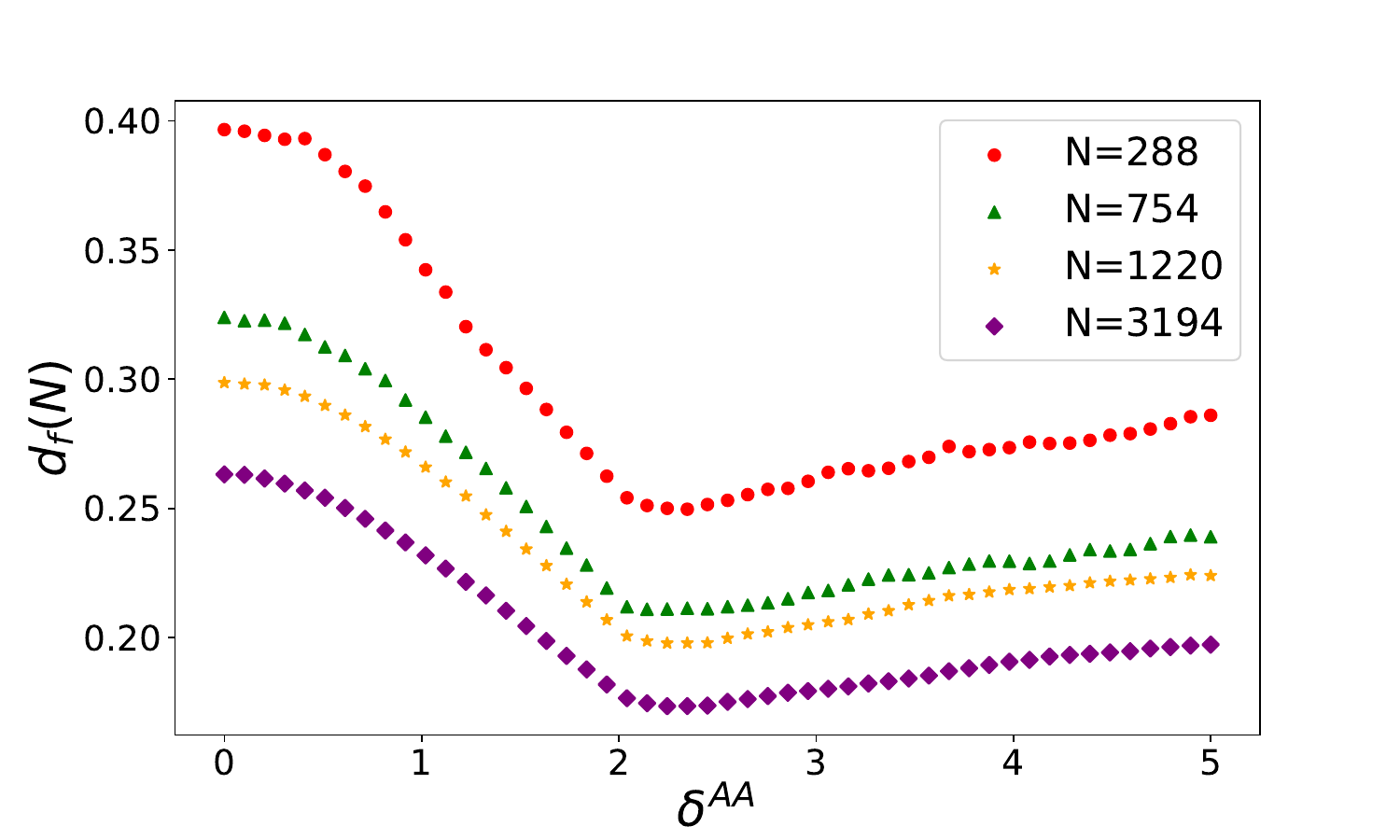}
    \includegraphics[keepaspectratio, width=0.45\textwidth]{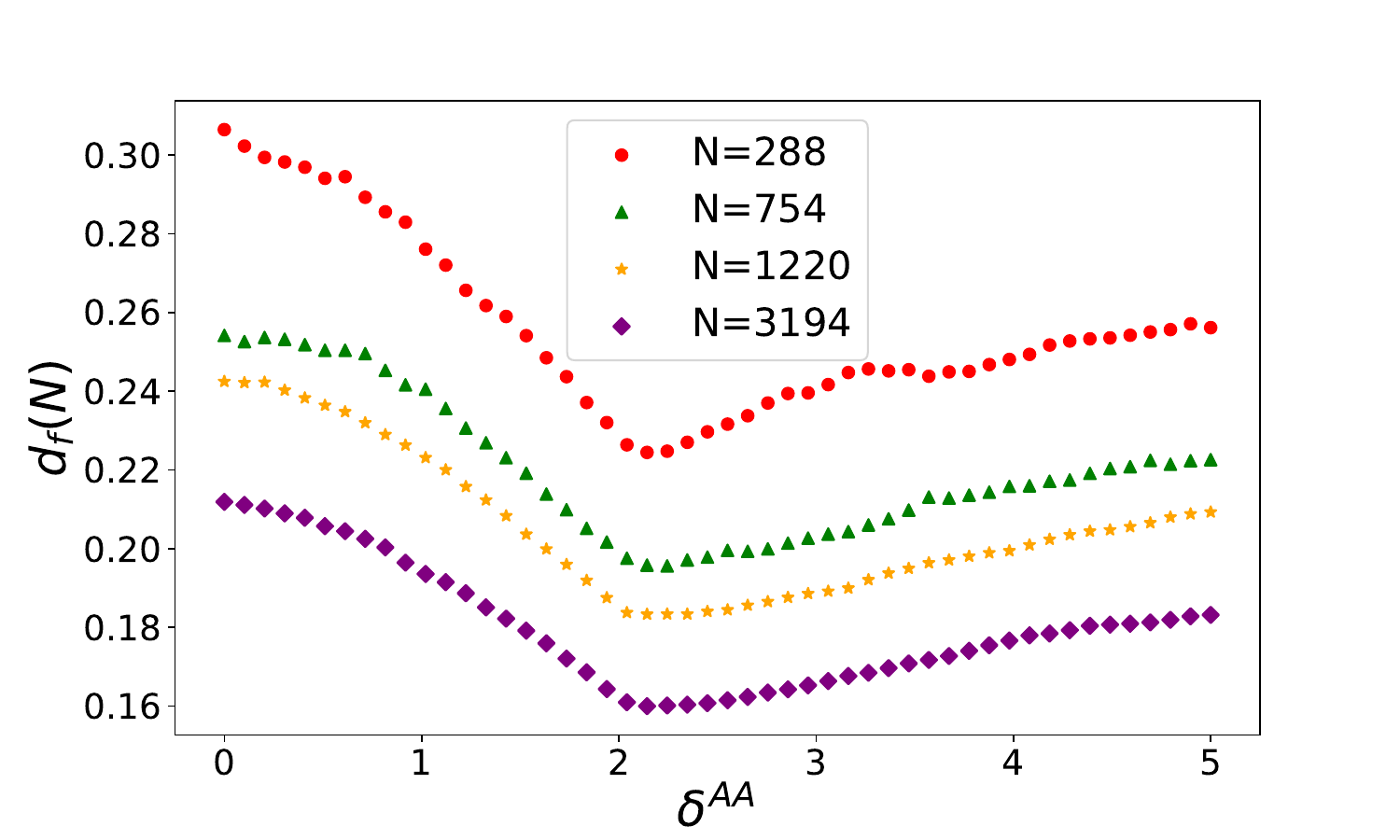}
    \put(-400,110){\textbf{(c) }}
    \put(-160,110){\textbf{(d) }}
    \\
    \caption{Plot of fractal dimension as a function of $\delta^{AA}$ in real space for a fixed value $\delta_{d}$, $\phi=0$ in the incommensurate case for various system sizes, starting from $\delta_{d}=0.7$. The plot illustrates the variation in fractal dimension with increasing discrete disorder, commencing from the scenario depicted in Fig.~\ref{fig:combined_vector_plot3}. As the discrete disorder in intra-cellular spring stiffness escalates, the transition from the extended phase to the localized phase begins to exhibit bi-directionality in the parameter range $K_{a}>K_{b}$, accompanied by the presence of a complex localization properties. (a): $K_{a}<K_{b}$ in real space for $\delta_{d}=1.5$, (b): $K_{b}<K_{a}$ in real space for $\delta_{d}=0.7$, (c): $K_{b}<K_{a}$ in real space for $\delta_{d}=2$, (d): $K_{b}<K_{a}$ in real space for $\delta_{d}=3$. }
    \label{fig:combined_vector_plot6}
\end{figure*}

\begin{equation}\label{eq:19}
    u^{A}_{N}=(-1)^{N-1}\prod_{j=1}^{j=N-1}\frac{-(K_{a}+\delta^{d}_{j})}{K_{b}+\delta^{AA}\cos(2\pi \beta (j+1)+\phi)}u^{A}_{1}.
\end{equation}
From above we can calculate the relevant Lyapunov exponent $\gamma$, which is inverse of the localization length \cite{PhysRevLett.113.046802,SCALES199727}
\begin{equation}
    \gamma=-\lim\limits_{N \to \infty}\frac{1}{N}\ln\left|\frac{u^{A}_{N}}{u^{A}_{1}}\right|,
\end{equation}
which by using Eq.~\ref{eq:19} reduces to,
\begin{equation}\label{eq:lyapunon1}
    \gamma=-\lim\limits_{N \to \infty}\frac{1}{N}\sum_{j=1}^{N-1}\ln\left|\frac{K_{b}+\delta^{AA}\cos(2\pi\beta(j+1)+\phi)}{K_{a}+\delta^{d}_{j}}\right|.
\end{equation}
With $\delta_{d}$ set to zero and $AA$ modulation incommensurate with the underlying lattice, the summation above can be transformed into a Riemann integral. By using the properties of irrational rotation \cite{06bacaea-d5a8-3177-a6ba-b77181df9af3}, one can express it as an integral over the angular component $\theta\in[0,2\pi]$
\begin{equation}\label{eq:integral1}
    \gamma=\frac{1}{2\pi}\int_{0}^{2\pi}\ln\left|\frac{K_{b}+\delta^{AA}\cos(\theta)}{K_{a}}\right|d\theta.
\end{equation}
The Eq.~\eqref{eq:integral1} can be evaluated to 
\begin{equation}
    \gamma = \ln\left(\frac{K_{b}+\sqrt{K_{b}^{2}-(\delta^{AA})^{2}}}{2K_{a}}\right),
\end{equation}
for $K_{b}>\delta^{AA}$.
And for $K_{b}<\delta^{AA}$ the expression be 
\begin{equation}
    \gamma=\ln\left(\frac{\delta^{AA}}{2K_{a}}\right).
\end{equation}
Setting the Lyapunov exponent to $0$ calculates the boundary for topological phase transition, corresponding to a divergent localization length. We do not need to worry about the initial phase $\phi$, as its information dissipates during integration in incommensurate modulation. With $\delta_{d}=0$ and incommensurate modulation, the localization length at the central frequency can be numerically computed using the transfer matrix method \cite{SCALES199727} based on relevant parameters ($K_{a}$, $K_{b}$, $\delta_{d}$, $\delta^{AA}$). 
\subsection{Commensurate modulation}
\begin{figure*}[htbp]
    \centering
    \includegraphics[keepaspectratio, width=0.45\textwidth]{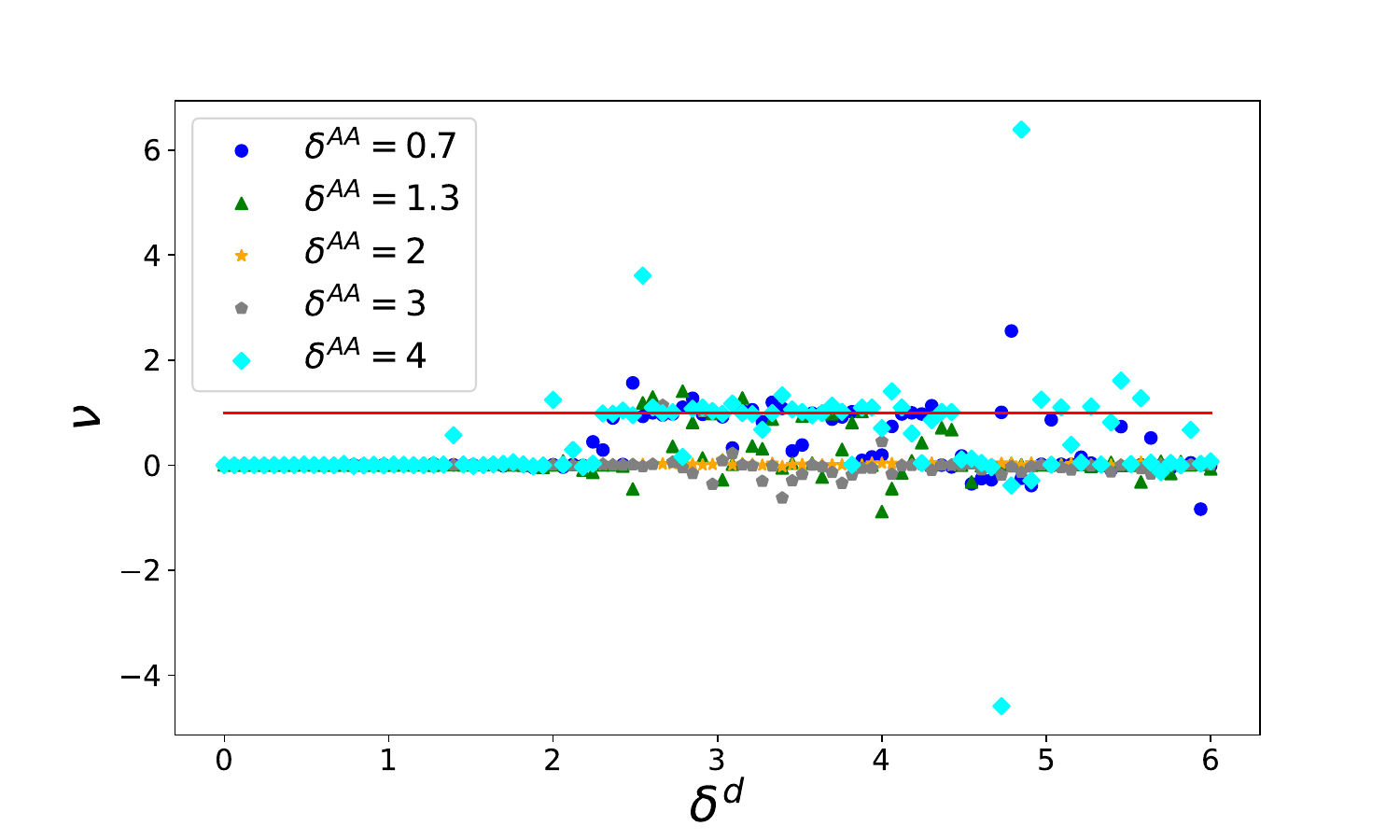}
    \includegraphics[keepaspectratio, width=0.45\textwidth]{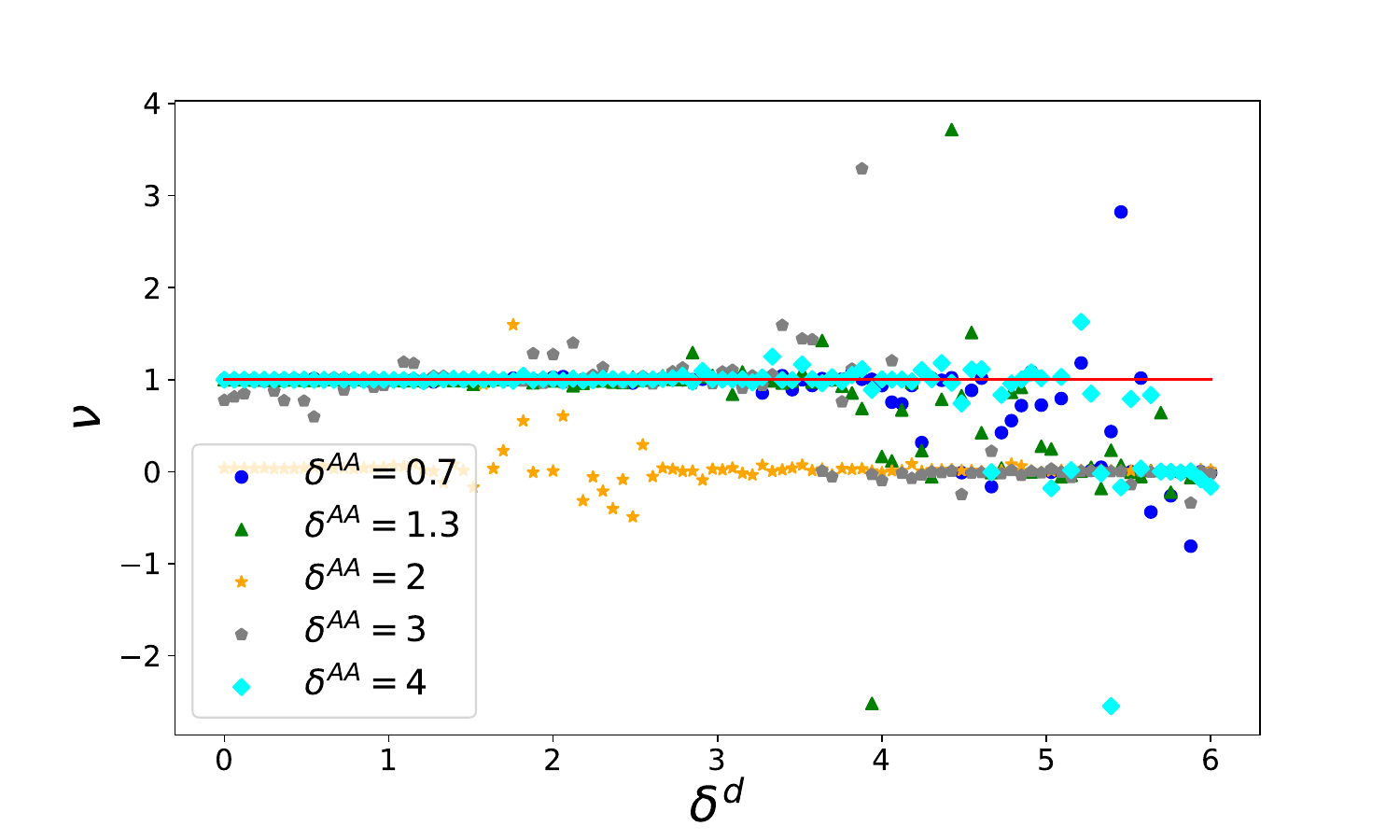}
     \put(-380,100){\textbf{(a) }}
     \put(-150,100){\textbf{(b)}}
     \\
     \caption{Evolution of LTM (topological invariant) plot for a single unit cell, away from the boundary with increasing discrete disorder. (a) System starts from a trivial phase $K_{a}>K_{b}$ with $\delta_{d}=0$, then as the disorder increases, the topological marker abruptly jumps to 1. Subsequently, with further increase in disorder strength, the topological marker drops back to 0, resulting in a return to a trivial phase. The fluctuation indicates that the topological invariant doesn't converge to $0$ or $1$ near the phase transition region. (b) System starts from a non-trivial phase in case of $K_{a}<K_{b}$. Please note that in the figure the axes labels must be $\delta_{d}$ rather than $\delta^{d}$.}
    \label{fig:ExampleAB}
\end{figure*}
This section builds on the method from Sec.~\ref{sec:b} to determine the phase transition boundary for commensurate modulation in topological phase transitions. In Eq.~\eqref{eq:19}, the expression cannot be directly converted into an integral due to the commensurate characteristics of the modulation. The initial step is to make the substitution, $\beta=\frac{s}{t}$, where $s$ and $t$ are coprime,
\begin{equation}
    \gamma=-\lim\limits_{N \to \infty}\frac{1}{N}\ln\left|\prod_{j=1}^{N-1}\frac{(K_{a}+\delta^{d}_{j})}{K_{b}+\delta^{AA}\cos(2\pi\frac{s}{t}(j+1)+\phi)}\right| .
\end{equation}
For simplicity $\theta_{j}=2\pi\frac{s}{t}(j+1)$.
Setting $\delta_{d}=0$ with specific language of modular mathematics, those terms of type $j+1=j+1 (\text{mod}t)$, will appear in the product, which implies that there will be $t$ different terms in the product, with each term appearing $\frac{N-1}{t}$ times. The resulting expression is 
\begin{equation}
    \gamma=-\lim\limits_{N \to \infty}\frac{1}{N}\ln\left|\prod_{j=1}^{j=t}\left(\frac{K_{a}}{K_{b}+\delta^{AA}\cos(\theta_{j}+\phi)}\right)\right| .
\end{equation}
In the large $N$ approximation the above expression can be simplified as 
\begin{equation}
    \gamma=\frac{1}{t}\ln\left|\prod_{j=1}^{t}\left(\frac{K_{b}+\delta^{AA}\cos(\theta_{j}+\phi)}{K_{a}}\right)\right| .
\end{equation}
In a way similar to the incommensurate scenario, the boundary of the topological phase transition can be identified by putting the value of the Lyapunov exponent to 0, which results in divergence in the localization length. When the periodicity limit of the modulation is small, i.e., when $\frac{sN}{t}\ll1$, the transition boundary in the commensurate situation can be determined as

\begin{equation}
    K_{a} = e^{Ti\pi}(K_{b} + \delta^{AA}\cos(\phi))
\end{equation}

for both even and odd values of $T$. In this scenario, the topological phase transition exhibits rich phase characteristics as shown in Fig.~\ref{fig:vector_plot91} and Fig.~\ref{fig:vector_plot92}, hence manipulating 1D edge modes becomes more tricky due to the additional influence of the modulation phase shift $\phi$.

\subsection{Inverse participation ratio and finite-size scaling}
\begin{figure*}[htbp]
    \centering
    \includegraphics[keepaspectratio, width=0.45\textwidth]{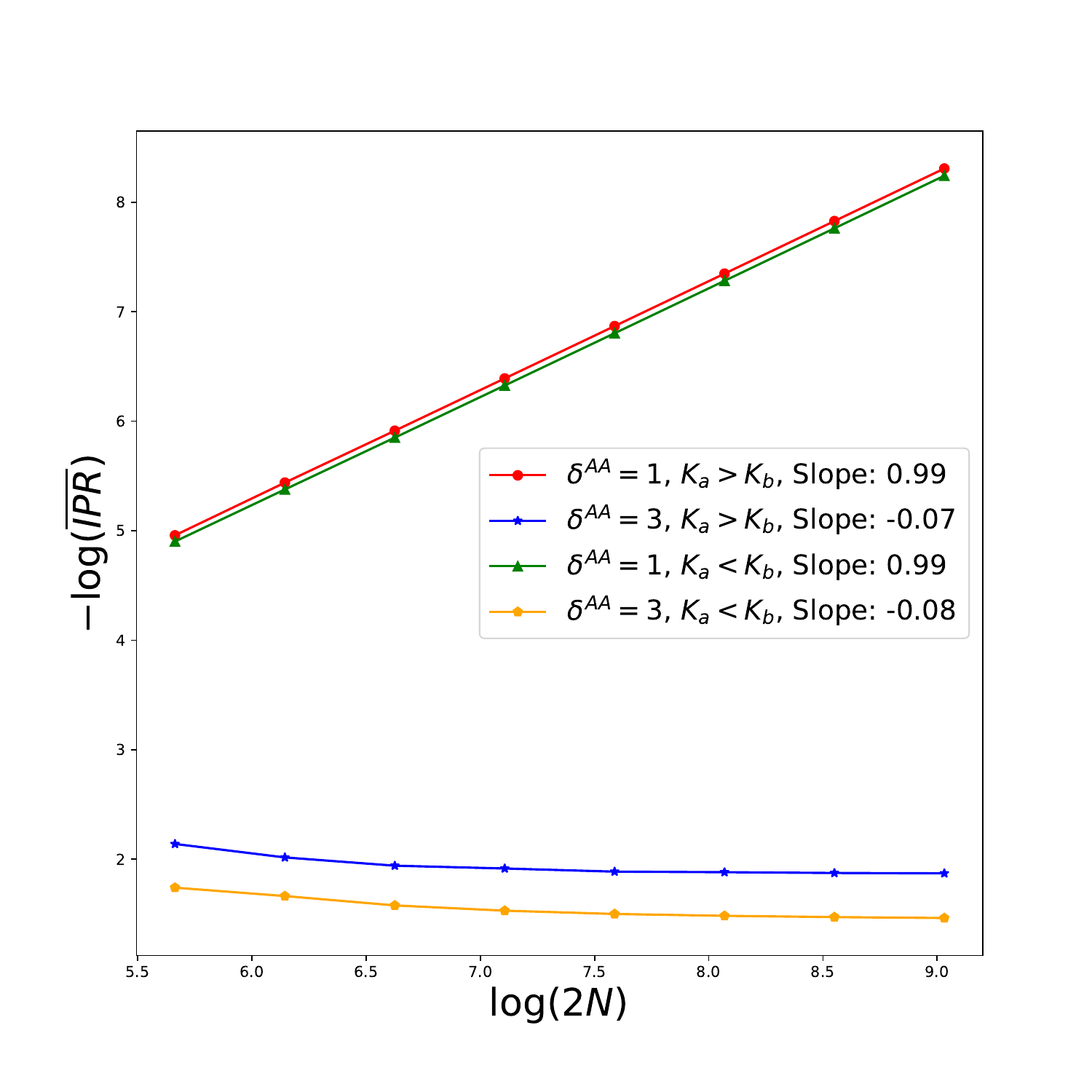}
    \includegraphics[keepaspectratio, width=0.45\textwidth]{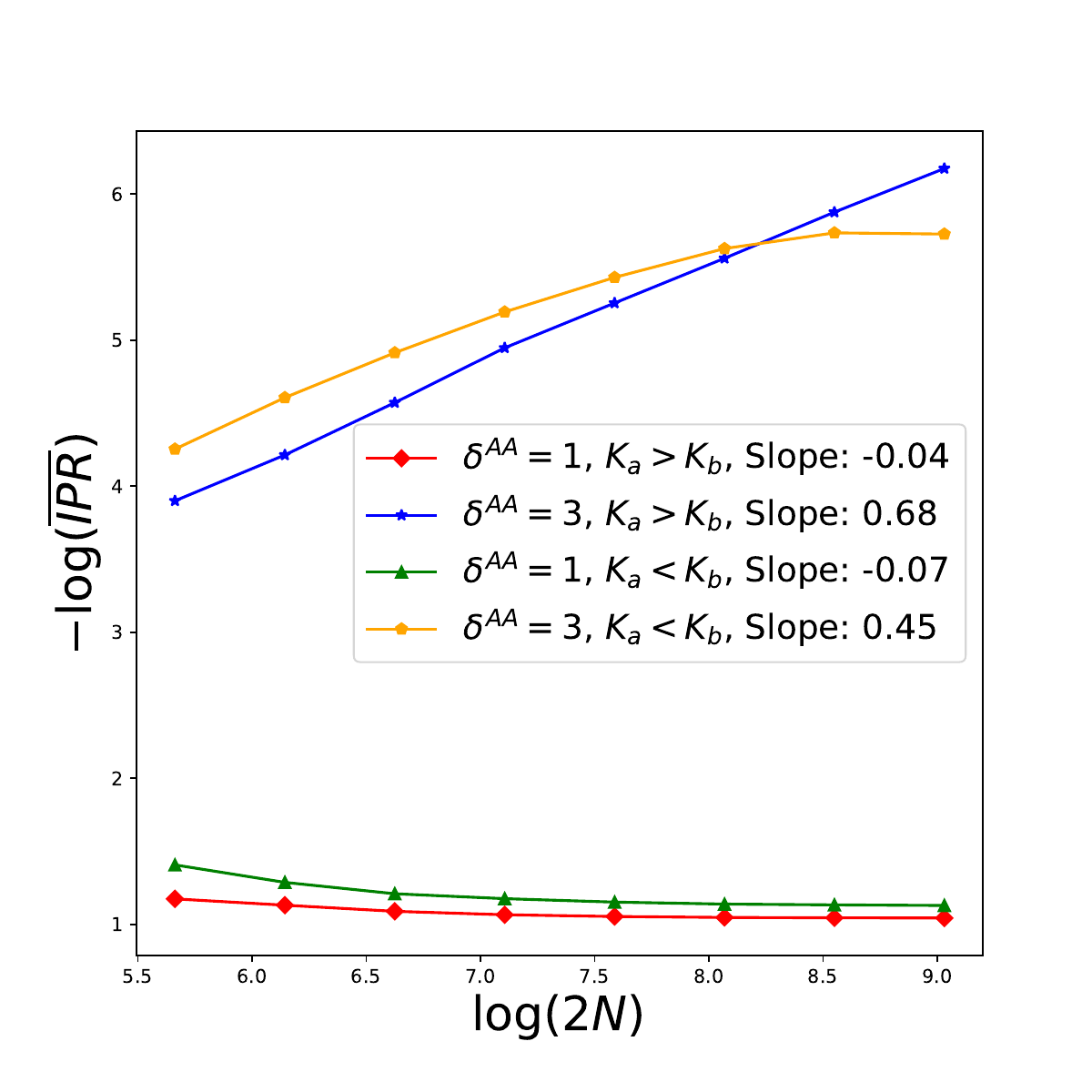}
     \put(-380,70){\textbf{(a) }}
     \put(-150,70){\textbf{(b)}}
     \\
    \caption{Plot of $-\ln(\overline{IPR})$ versus $\ln(2N)$ for $\delta_{d}=0$ and $\phi=0$, for the following cases. The slope of the curve indicates the value of $d_{f}(N)$ which, as $N$ becomes large, converges to the fractal dimension. The upper curve is for $\delta^{AA}=1$ and the lower curve is for $\delta^{AA}=3$. Case (a) represents real space, and case (b) represents momentum space, respectively. In case (a), the red and green curves ($\delta^{AA}<2$) indicate that the system is in the extended phase, whereas the blue and orange curves ($\delta^{AA}>2$) indicate that the system is in the localized phase, as expected. In case (b), the red and green curves indicate the localized phase, whereas the orange curves indicate the extended phase. The blue curve in case (b) is particularly interesting; here, $d_{f}(N)$ does not converge to 0 or 1 as shown by the blue curve slope. Therefore, one can quantitatively estimate the presence of a critical phase here.}
    \label{fig:combined_vector_plot4}
\end{figure*}

Analyzing the inverse-participation ratio (IPR) enhances understanding of the localization properties of the model. The IPR for a given eigenstate is defined as 
\begin{equation}
    IPR(|\psi\rangle)=\sum_{j=1}^{2N}|u_{j}|^{4}
\end{equation}
where $u_{j}$ are the components of the normalized eigenstate $|\psi\rangle$, which has a dimension of $2N$.
For an extended eigenmode, the IPR scales as $\frac{1}{N}$, with $N$ representing the system size, while a fully localized eigenmode has an IPR that scales as $1$. Eigenmodes that do not fit these classifications indicate the critical phase of the phase transition \cite{RevModPhys.80.1355,PhysRevLett.126.106803,PhysRevB.105.174206,PhysRevB.109.014210}. The IPR of eigenstates can be plotted with energy eigenvalues against $A-A$ strength $\delta^{AA}$ at a fixed $\delta_{d}$, or against $\delta_{d}$ with a fixed $\delta^{AA}$. These plots effectively illustrate the transition among localized, de-localized, and critical phases in both real and momentum spaces.
In studying localization with $AA$ modulation, a crucial parameter to calculate is fractal dimension $D_{f}$. This dimension is independent of the size of the system and dictates the long-term behavior of the IPR (or the average IPR) as the system size $N$ changes. The scaling relationship is given by $\overline{IPR} \approx (2N)^{-D_{f}}$.
Accessing $D_{f}$ from values of systems having finite sizes is achievable through the use of a quantity $d_{f}(N)$, that depends on the system and which leads to fractal dimension $D_{f}$ as $N$ grows large. The expression for $d_{f}(N)$ is defined as:
\begin{equation}
    d_{f}(N)=-\frac{\ln(\overline{IPR})}{\ln(2N)}.
\end{equation}
In the context where $2N$ denotes the system size, the fractal dimension $D_{f}$ equals $1$ in the delocalized phase and $0$ in the localized phase \cite{https://doi.org/10.1002/andp.202200203}. Numerical computations show that as $N$ increases, $d_{f}(N)$ approaches $D_{f}$. When $\delta^{AA}$ is fixed and $\delta_{d}$ varies, or vice versa, $d_{f}(N)$ trends towards $1$ in the delocalized phase and $0$ in the localized region (while a finite portion of states remains localized). In the critical region, $d_{f}(N)$ stabilizes at a finite value between 0 and 1, independent of $N$.
\begin{figure*}[htbp]
  \centering
  \begin{subfigure}[b]{0.4\textwidth}
    \centering
    \includegraphics[width=\textwidth]{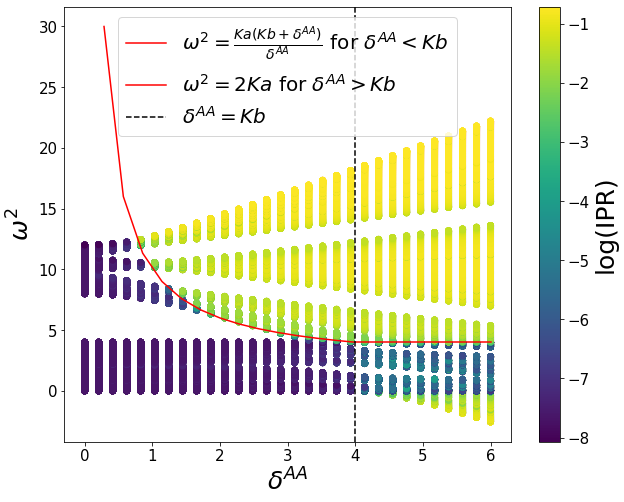}
    \caption{$K_{a}<K_{b}$ and $\delta_{d}=0$}
    \label{fig:nonchiral1}
  \end{subfigure}
  \begin{subfigure}[b]{0.4\textwidth}
    \centering
    \includegraphics[width=\textwidth]{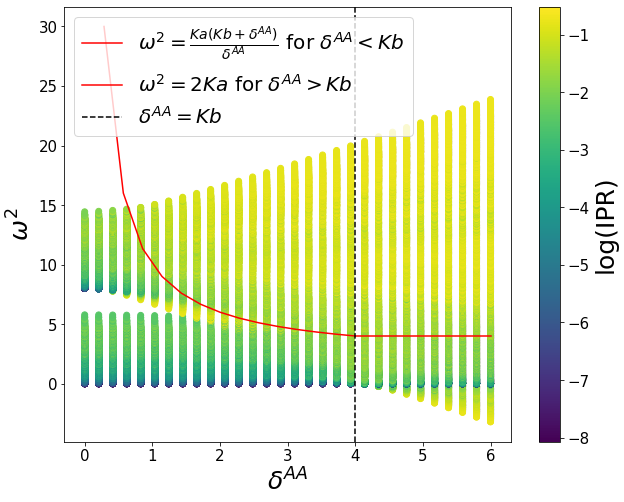}
    \caption{$K_{a}<K_{b}$ and $\delta_{d}=1.5$}
    \label{fig:nonchiral3}
  \end{subfigure}
  \hspace{0.05\textwidth}
  \begin{subfigure}[b]{0.4\textwidth}
    \centering
    \includegraphics[width=\textwidth]{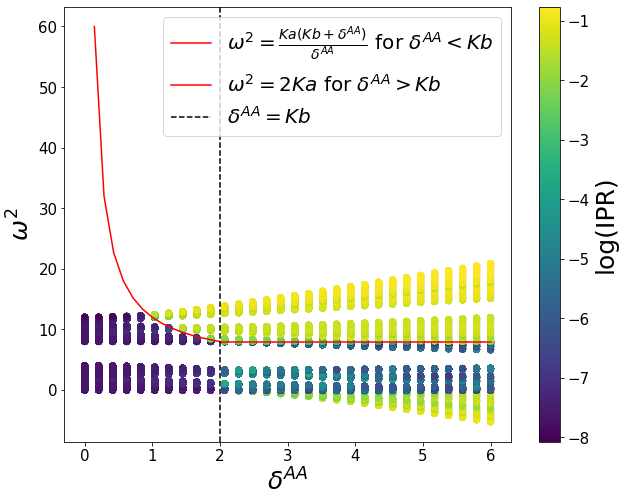}
    \caption{$K_{a}>K_{b}$ and $\delta_{d}=0$}
    \label{fig:nonchiral2}
  \end{subfigure}
  \begin{subfigure}[b]{0.4\textwidth}
    \centering
    \includegraphics[width=\textwidth]{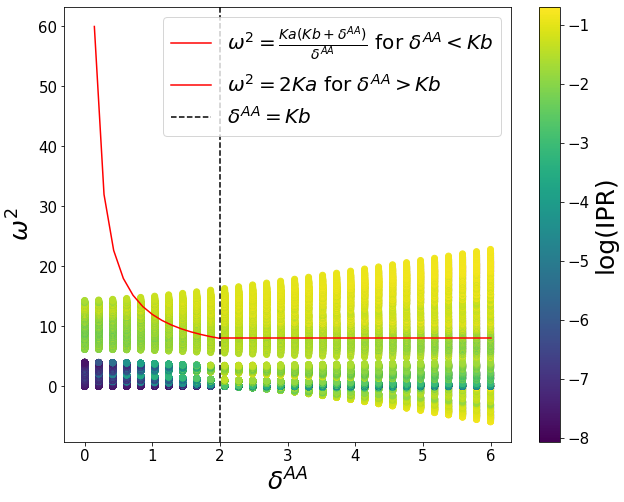}
    \caption{$K_{a}>K_{b}$ and $\delta_{d}=1.5$}
    \label{fig:nonchiral4}
  \end{subfigure}
  \caption{Plot of IPR as function of eigenvalues of the actual non-chiral model for $N=1597$ unit cells, $\phi=0$. The red line indicates the result obtained in Eq.~\eqref{eq:F1}, Eq.~\eqref{eq:F2}, the boundary between localized and non-localized eigenmodes, without discrete disorder. The black splitted line indicates $\delta^{AA}=K_{b}$ as shown in original phase plot. The Fig.~\ref{fig:nonchiral3} and Fig.~\ref{fig:nonchiral4} as shown is for $\delta_{d}=1.5$. In cases (b) and (d), critical states appear below the critical strength of quasi-periodic modulation $\delta^{AA}=2$ with a new analytic expression for the mobility edge, which cannot be accounted for, using Ref.~\cite{10.1007/s11511-015-0128-7}.}
  \label{fig:Nonchiral}
\end{figure*}

\section{Calculation of mobility edges without disorder in intra-cellular spring stiffness}\label{sec:4}
The mobility edge is a key characteristic of quasi-periodic systems, extensively examined in the literature. It represents the boundary between extended and localized eigenstates. In one dimension, mobility edges are less likely to occur in quasi-periodic systems compared to Anderson like disorder in first two spatial dimensions. This study shows that mobility edges can be analytically derived in the original non-chiral version of the system, whereas the chiral model removes eigenvalue dependence due to an additional on-site spring.
Setting $\delta_{d}=0$, the equation of the motion in non-chiral representation 
    \begin{equation}
        \begin{aligned}
\omega^{2}u^{A}_{j} &= (K_{a} + K_{b} + \delta^{AA}\cos(2\pi \beta j+\phi))u^{A}_{j} \\
                     &\quad -(K_{a})u^{B}_{j} \\
                     &\quad -(K_{b} + \delta^{AA}\cos(2\pi \beta j + \phi))u^{B}_{j-1},
\end{aligned}
    \end{equation}
    \begin{equation}
        \begin{aligned}
\omega^{2}u^{B}_{j} &= (K_{a} + K_{b} + \delta^{AA}\cos(2\pi \beta (j+1)+\phi))u^{B}_{j} \\
                     &\quad -(K_{a})u^{A}_{j} \\
                     &\quad -(K_{b} + \delta^{AA}\cos(2\pi \beta (j+1) + \phi))u^{A}_{j+1}.
\end{aligned}
    \end{equation}
The above equations can be solved for 
\begin{equation}
    u^{B}_{j}=\frac{K_{a}u^{A}_{j}+(K_{b}+\delta^{AA}\cos(2\pi \beta(j+1)+\phi))u^{A}_{j+1}}{K_{a}+K_{b}-m\omega^{2}+\delta^{AA}\cos(2\pi \beta(j+1)+\phi)}.
\end{equation}

Using the expression derived in Section.~\ref{subsec:2}, Eq.~\eqref{eq:f}, replacing only $\Delta_{1}=\delta^{AA}$ we have  
\begin{equation}\label{eq:F1}
    m\omega^{2}=2K_{a}
\end{equation} for $K_{b}<\delta^{AA}$
\begin{equation}\label{eq:F2}
    m\omega^{2}=\frac{K_{a}(K_{b}+\delta^{AA})}{\delta^{AA}}
\end{equation} for $K_{b}>\delta^{AA}$.
The analytical result above is compared with Fig.~\ref{fig:Nonchiral}. It can be observed that all three localization regimes coexist within our calculated regions as shown in Fig.~\ref{fig:nonchiral1} and Fig.~\ref{fig:nonchiral2}. All eigenstates above the red boundary are localized, and below, there exist two possible regions: for $K_{b}<\delta^{AA}$, all states are in the critical regime, whereas for $K_{b}>\delta^{AA}$, all states are in the extended eigenmode. Therefore, the critical regime is distinctly defined with boundaries given by $m\omega^{2}=2K_{a}$ and $K_{b}<\delta^{AA}$. As the discrete disorder is increased, the mobility edge no longer remains valid, and the states become more and more critical in nature as shown in Fig.~\ref{fig:nonchiral3} and Fig.~\ref{fig:nonchiral4}.
\section{Conclusion}
In this study we conducted a detailed exploration of the topological phase transition in a mechanical SSH model featuring Andre-Aubry periodic modulation of spring stiffness and discrete disorder. We maintained the chiral symmetry of the dynamical matrix throughout our computations. Consequently, we employed a topological invariant from class-AIII or DBI, along with its real space covariant form, to analyze and characterize the topological phases. \par

We derived an analytical expression for the Lyapunov exponent without discrete disorder, which enables prediction of the topological phase transition boundary based on system parameters ($K_{a}$, $K_{b}$, $\delta_{d}$, $\delta^{AA}$). We examined the localization properties of the eigenmodes by calculating the state IPR and fractal dimensions, categorizing the entire eigenspectrum into extended, non-localized, and critical regimes. The fractal dimension calculations facilitated a comprehensive analysis of the localization properties and supported finite-size scaling analysis. \par
We calculated the boundary between localized and non-localized regimes in the non-chiral model, while also restoring the eigenvalue dependence of the diagonal terms. In the absence of discrete disorder in intra-cellular spring stiffness, well-defined boundaries for critical, localized, and extended states emerged. However, as discrete disorder increased, the mobility edge diminished and states began to mix. The IPR calculation revealed a TAI phase, where eigenstates started to localize after reaching a certain level of disorder, accompanied by critical states. The evolution of the topological invariant showed a similar trend. Additionally, background quasi-periodic modulation adjusted the topological phase transition boundary, influencing the localization properties of the eigenmodes.

Finally, during the submission of this article, the author noticed a recently published article entitled- `Mechanical Su-Schrieffer-Heeger quasicrystal: Topology, localization, and mobility edge' \cite{PhysRevB.109.195427}, in which a similar type of problem has also been studied, taking into account only the effect of quasi-periodic modulation.

\bibliographystyle{apsrev4-2}
\bibliography{bibliography3}

\appendix

\section{$A-A$ quasi-periodic modulation in 1D spring-mass stiffness constants(both intracellular and intercellular modulation)}
The equation of motions in the presence of two quasi-periodic $AA$ modulations are 
\begin{widetext}
\begin{equation}
m\ddot{u}_{A/B}^{j} = -(K_{a} + K_{b} + \Delta_{1}\cos(2\pi Q_{1}(j/j+1) + \phi_{1}) + \Delta_{2}\cos(2\pi Q_{2}j + \phi_{2}))u_{A/B}^{j} + (K_{a} + \Delta_{2}\cos(2\pi Q_{2}j + \phi_{2}))u_{B/A}^{j} + Y,
\end{equation}
\end{widetext}
where $Y=(K_{b} + \Delta_{1}\cos(2\pi Q_{1}(j/j+1) + \phi_{1}))u_{B/A}^{(j-1)/(j+1)}$.

\begin{figure*}[htbp]
    \centering
    \includegraphics[keepaspectratio, width=0.45\textwidth]{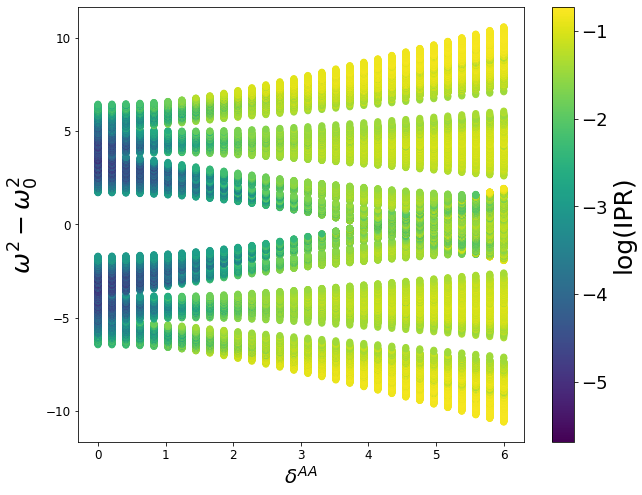}
    \includegraphics[keepaspectratio, width=0.45\textwidth]{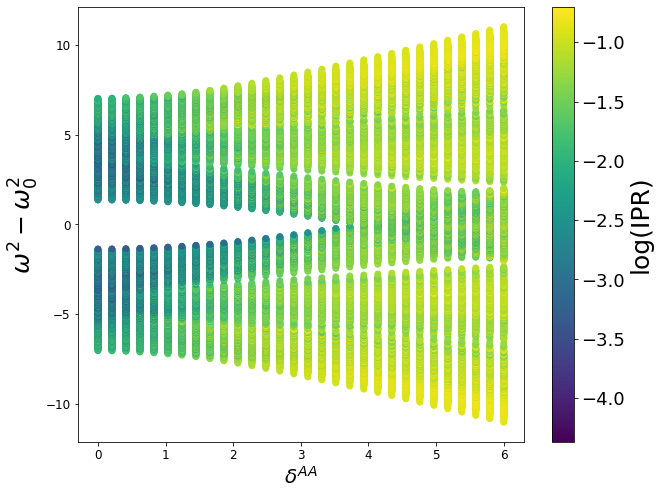}
     \put(-402,160){\textbf{(a) }}
     \put(-170,157){\textbf{(b)}}
     \\
    \includegraphics[keepaspectratio, width=0.45\textwidth]{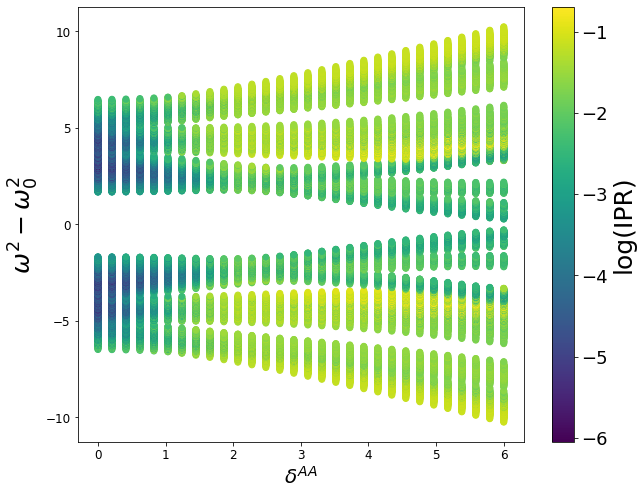}
    \includegraphics[keepaspectratio, width=0.45\textwidth]{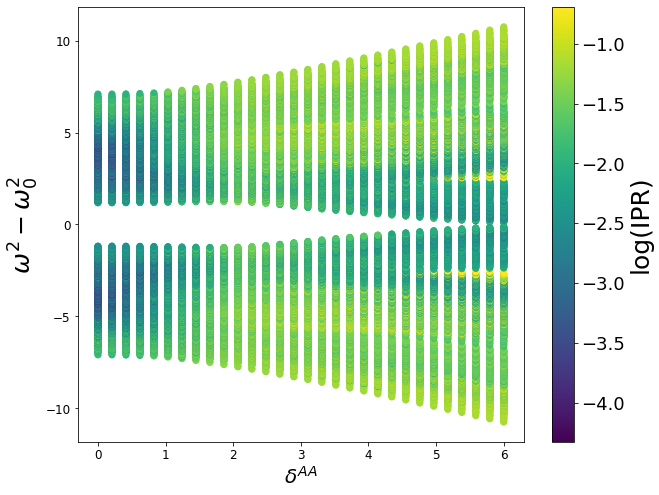}
    \put(-400,160){\textbf{(c) }}
    \put(-173,157){\textbf{(d) }}
    \\
    \caption{Plot of IPR versus $\delta^{AA}$ as function of eigenvalue spectrum $\omega^{2}-\omega_{o}^{2}$ for $N=1597$ unit cells and $\phi=0$ with fixed values of $\delta_{d}$. The plot details the change in the system localization spectrum starting with increasing discrete disorder $\delta_{d}$ starting from $\delta_{d}=0$ as shown in Fig.~\ref{fig:combined_vector_plot2}. (a): $K_{b}>K_{a}$ in real space for $\delta_{d}=0.7$, (b): $K_{b}>K_{a}$ in real space for $\delta_{d}=1.5$, (c): $K_{b}<K_{a}$ in real space for $\delta_{d}=0.7$, (d): $K_{b}<K_{a}$ in real space for $\delta_{d}=1.5$.}
    \label{fig:mainA}
\end{figure*}

\begin{figure*}[htbp]
    \centering
    \includegraphics[keepaspectratio, width=0.45\textwidth]{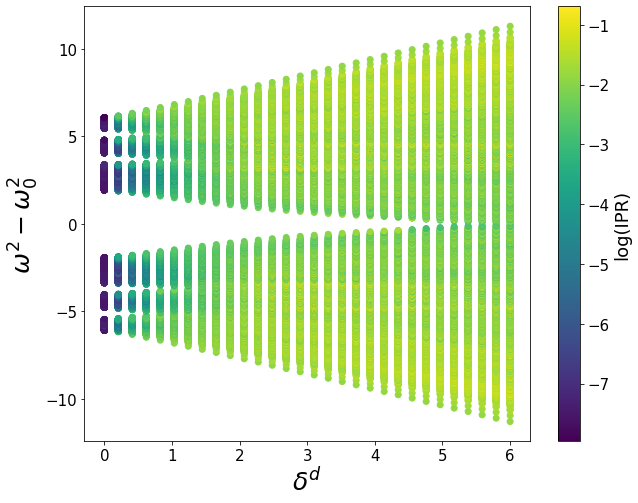}
    \includegraphics[keepaspectratio, width=0.45\textwidth]{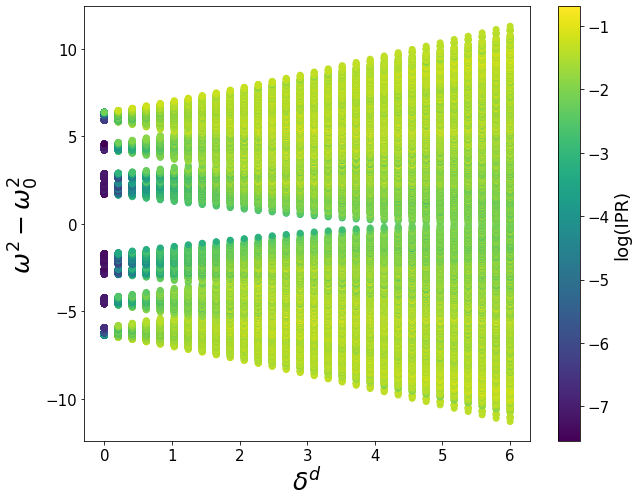}
     \put(-402,160){\textbf{(a) }}
     \put(-170,160){\textbf{(b)}}
     \\
    \includegraphics[keepaspectratio, width=0.45\textwidth]{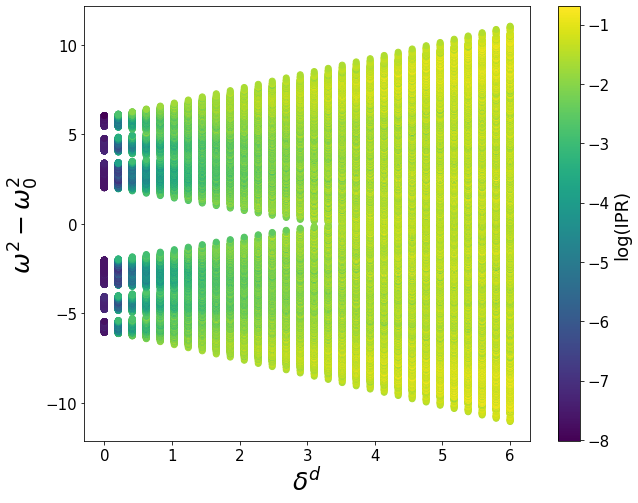}
    \includegraphics[keepaspectratio, width=0.45\textwidth]{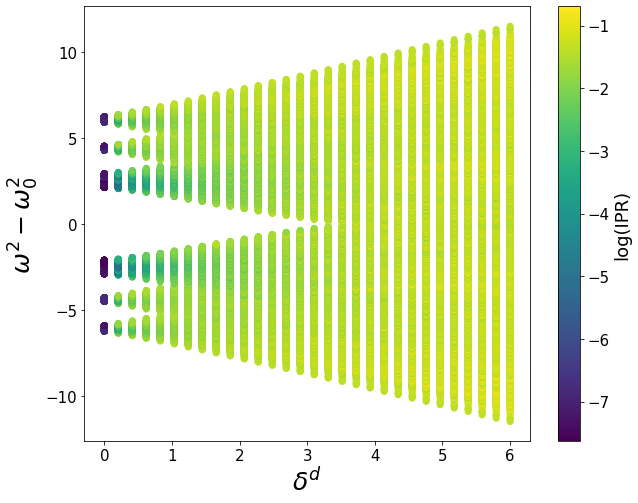}
    \put(-400,160){\textbf{(c) }}
    \put(-180,160){\textbf{(d) }}
    \\
    \caption{Plot of IPR versus $\delta_{d}$ in the chiral version of the model. The plots details the change in state IPR, hence localization properties with changing discrete disorder, $\delta_{d}$, for fixed value of quasi-periodic modulation $\delta^{AA}$, starting from $\delta^{AA}=1$. The plots where $K_{a}>K_{b}$ indicates localization of the eigenstate with increasing disorder in an intermediate regime, indicating a behaviour of Anderson insulator. (a):$K_{b}>K_{a}$ in real space for $\delta^{AA}=0.7$, (b):$K_{b}>K_{a}$ in real space for $\delta^{AA}=1.5$, (c):$K_{b}<K_{a}$ in real space for $\delta^{AA}=0.7$, (d):$K_{b}<K_{a}$ in real space for $\delta^{AA}=1.5$. Please note that in the figure the axes labels must be $\delta_{d}$ rather than $\delta^{d}$.}
    \label{fig:mainB}
\end{figure*}

\begin{figure*}[htbp]
    \centering
    \includegraphics[keepaspectratio, width=0.45\textwidth]{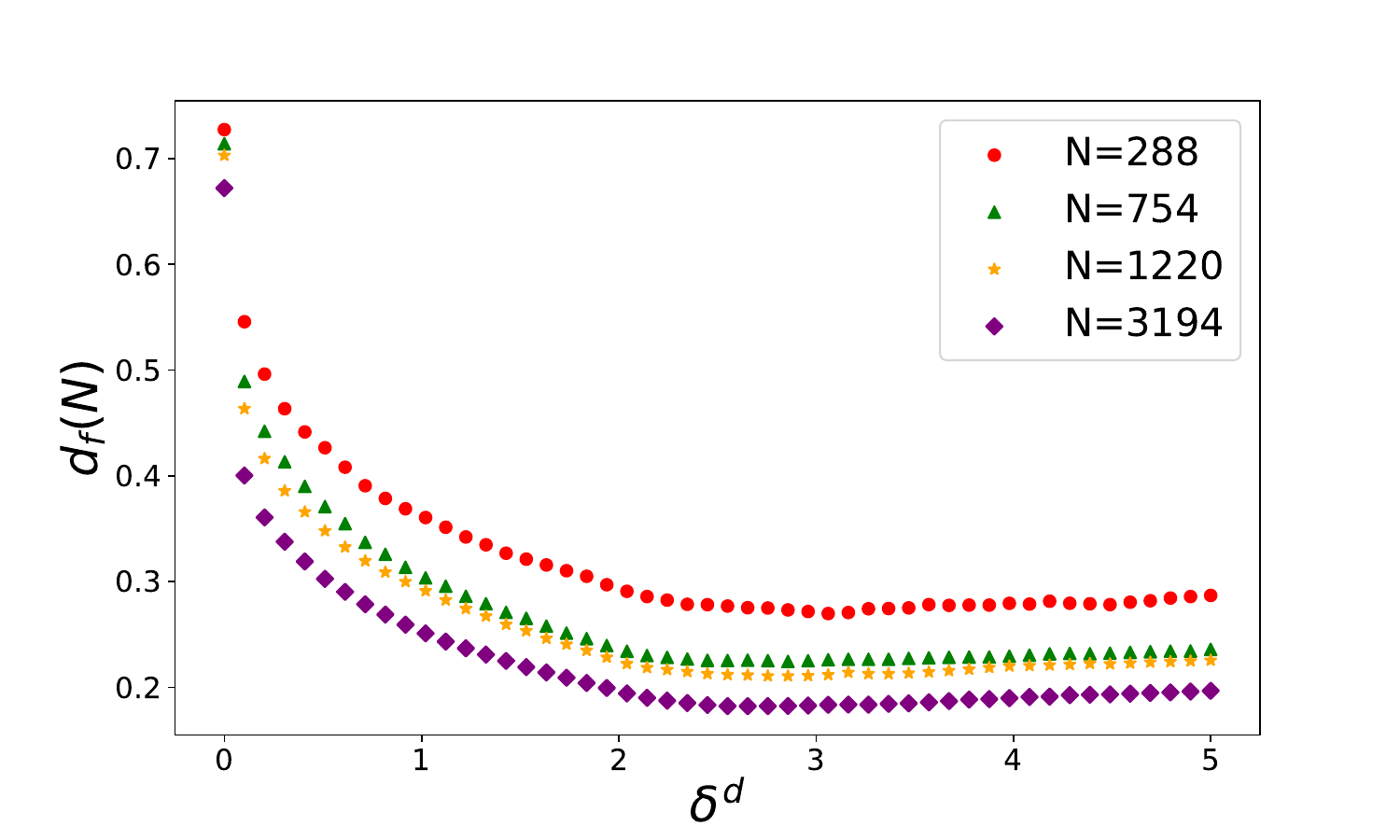}
    \includegraphics[keepaspectratio, width=0.45\textwidth]{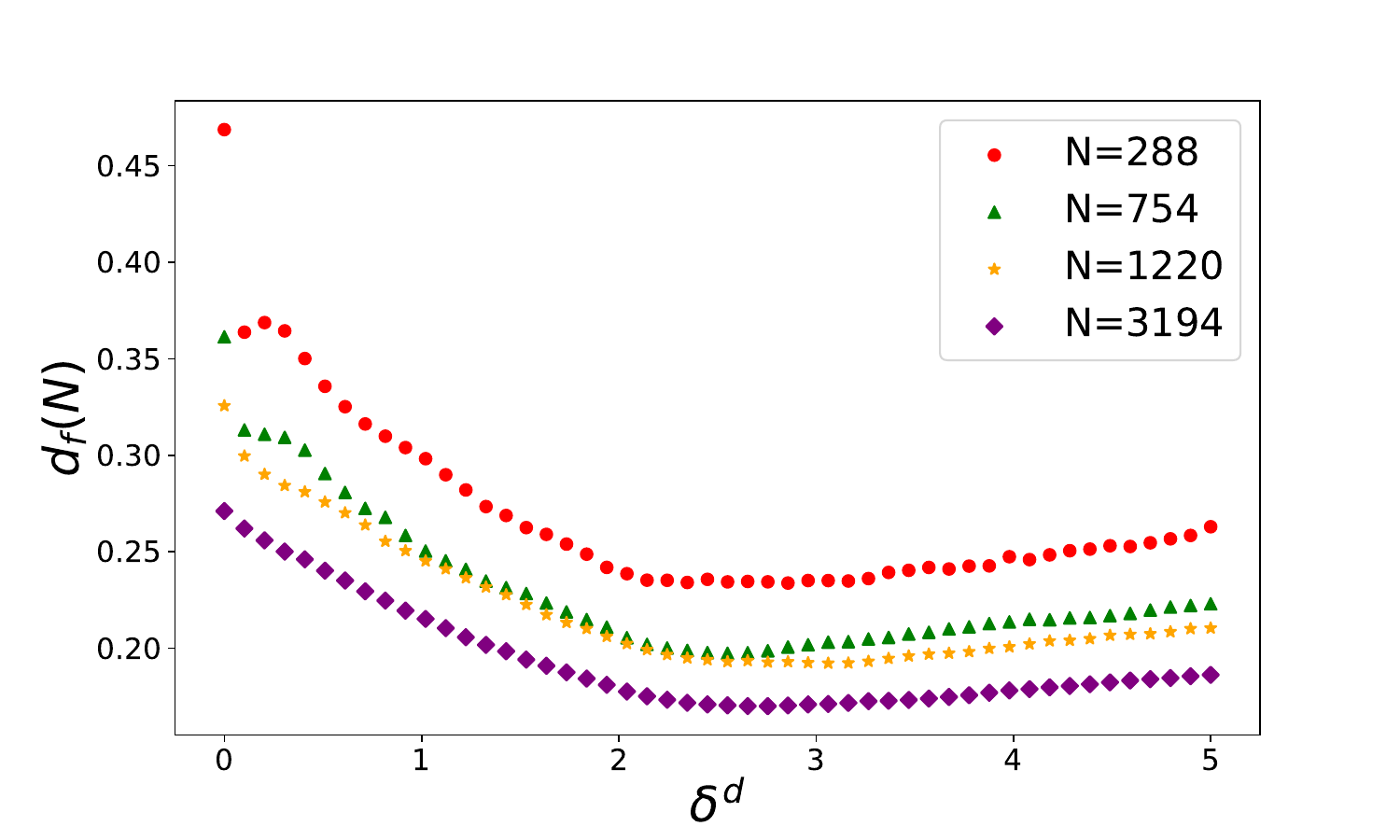}
     \put(-402,110){\textbf{(a) }}
     \put(-170,110){\textbf{(b)}}
     \\
    \includegraphics[keepaspectratio, width=0.45\textwidth]{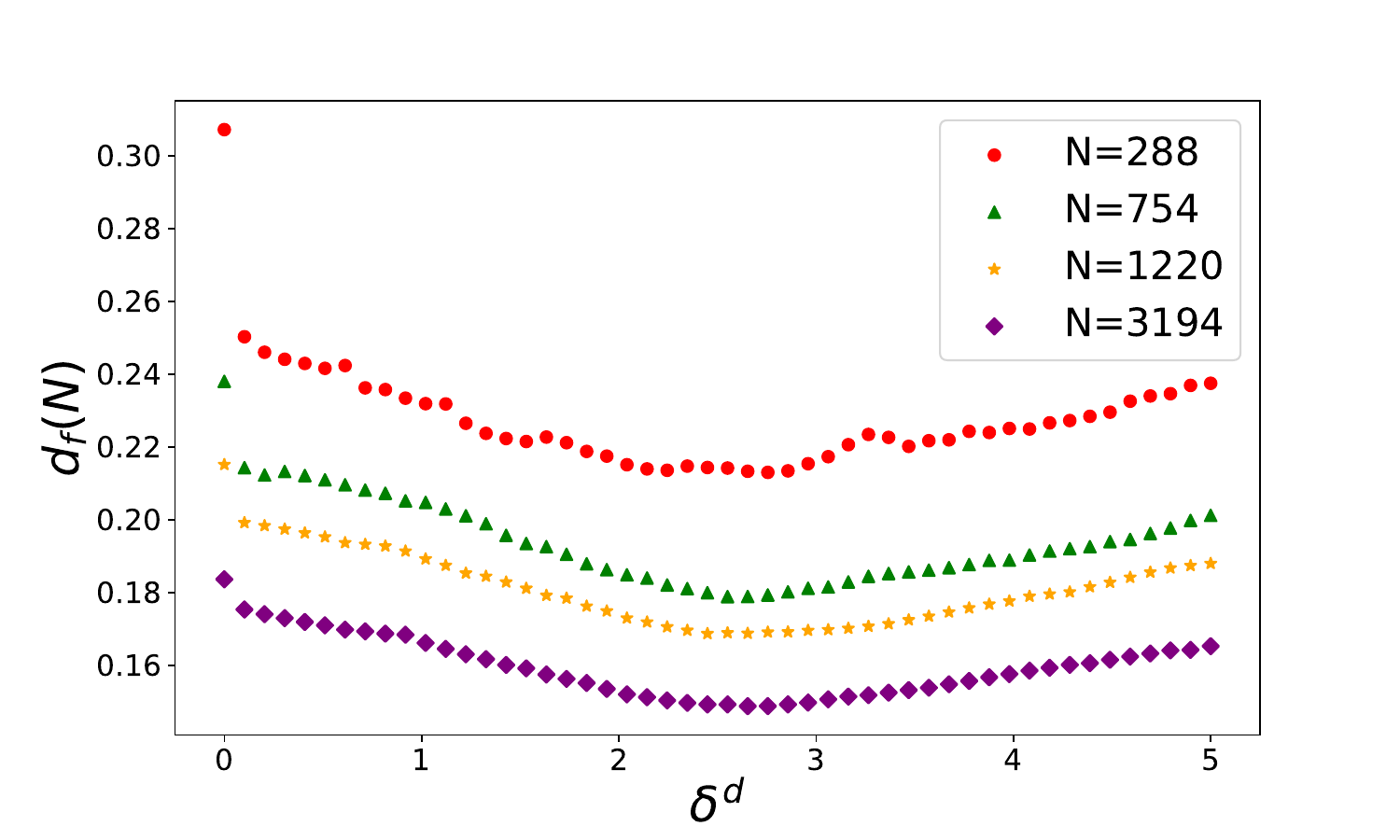}
    \includegraphics[keepaspectratio, width=0.45\textwidth]{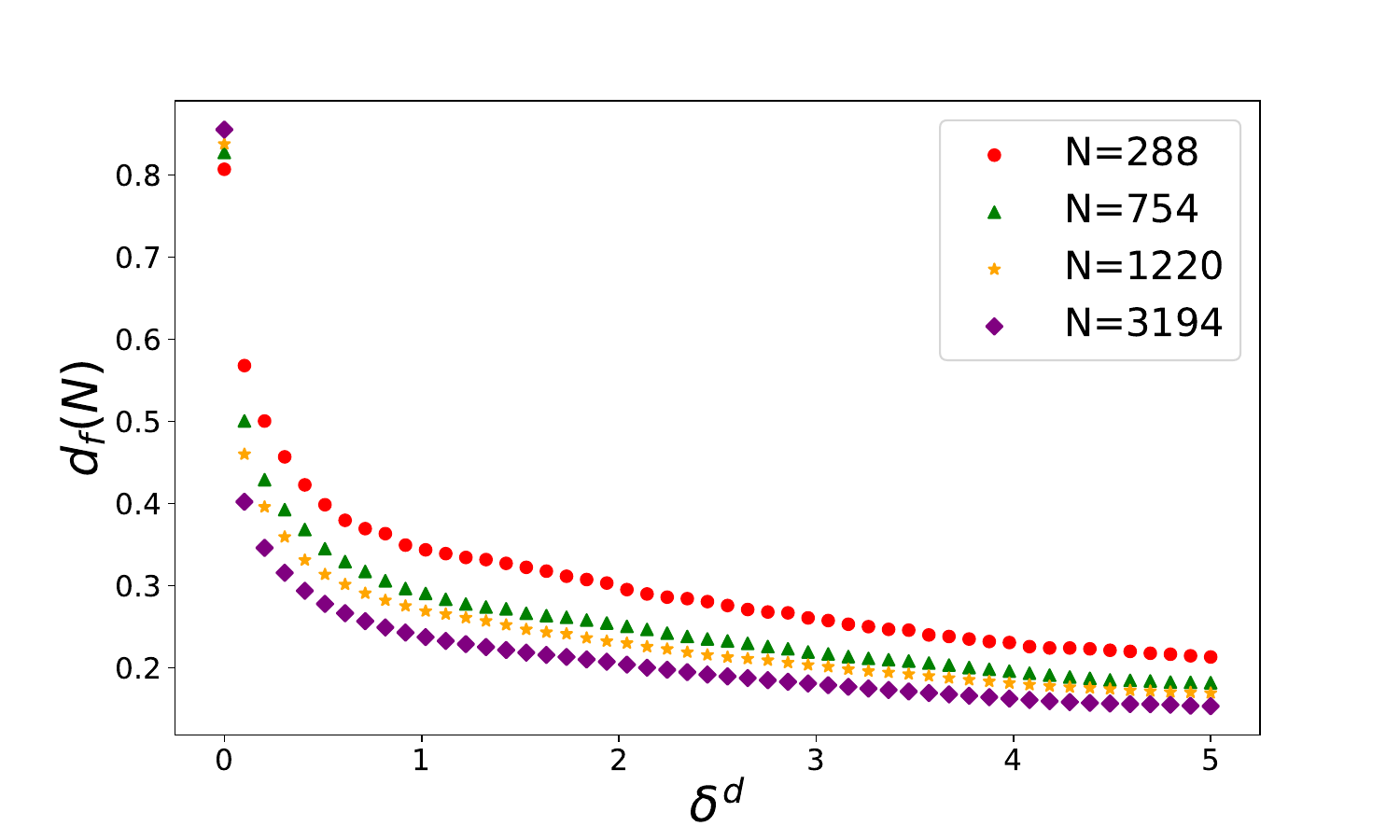}
    \put(-400,110){\textbf{(c) }}
    \put(-180,110){\textbf{(d) }}
    \\
    \includegraphics[keepaspectratio, width=0.45\textwidth]{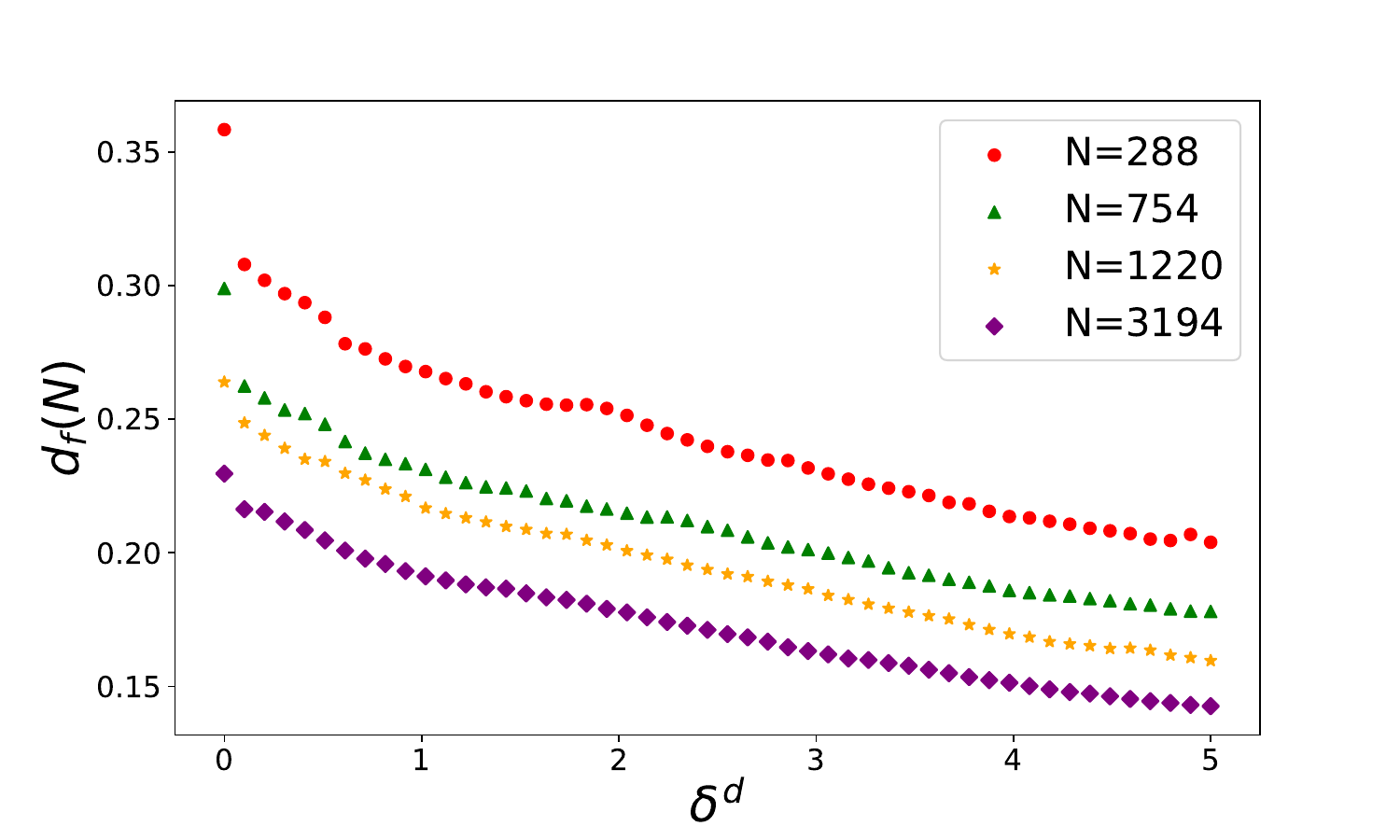}
    \includegraphics[keepaspectratio, width=0.45\textwidth]{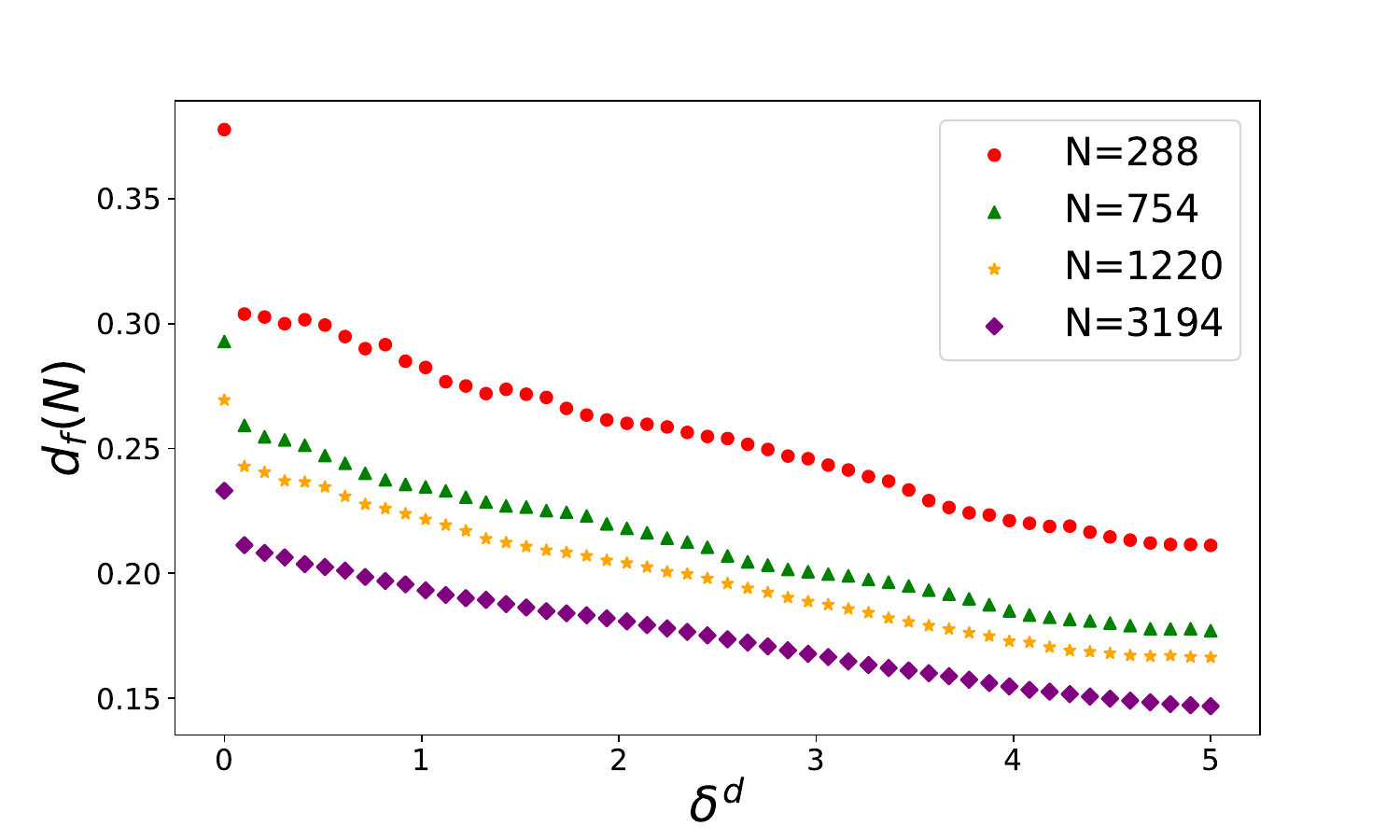}
    \put(-400,110){\textbf{(e) }}
    \put(-175,110){\textbf{(f) }}
    \caption{Plot of fractal dimension with discrete disorder for a fixed value of quasi-periodic modulation. As can be inferred from the plots in the case where $K_{a}<K_{b}$, with the increase of disorder the states are in critical phase as well. While for the case where $K_{a}>K_{b}$, the states are becoming localized. The plots of states IPR as a function energy also confirms the same. (a):$K_{a}<K_{b}$,  $\delta^{AA}=1.5$ , (b):$K_{a}<K_{b}$, $\delta^{AA}=2$, (c):$K_{a}<K_{b}$ and $\delta^{AA}=3$, (d):$K_{a}>K_{b}$ and $\delta^{AA}=1.5$, (e):$K_{a}>K_{b}$ and $\delta^{AA}=2$, (f):$K_{a}>K_{b}$ and $\delta^{AA}=3$. Please note that in the figure the axes labels must be $\delta_{d}$ rather than $\delta^{d}$.}
    \label{fig:combined_vector_plotA3} 
\end{figure*}

To maintain the chiral symmetry of the dynamical matrix, one must include an on-site stiffness constant in a particular manner,

\begin{equation}
    K_{0,A}^{j}=-\Delta_{1}\cos(2\pi Q_{1}j+\phi_{1})-\Delta_{2}\cos(2\pi Q_{2}j+\phi_{2}),
\end{equation}
\begin{equation}
    K_{0,B}^{j}=-\Delta_{1}\cos(2\pi Q_{1}(j+1)+\phi_{1})-\Delta_{2}\cos(2\pi Q_{2}j+\phi_{2}).
\end{equation}
So assuming the displacement satisfies 
\begin{equation}
    u_{\alpha}^{j}(t)=-\omega^{2}u_{\alpha}^{j}(t)
\end{equation}
where $\alpha={A,B}$, the dynamical matrix is 
\begin{widetext}
{\small
\begin{equation}
    D=\begin{bmatrix}
K_{a}+K_{b} & -(K_{a}+\Delta_{2}\cos(2\pi Q_{2}+\phi_{2})) & \dots & -(K_{b}+\Delta_{1}\cos(2\pi Q_{1}+\phi_{1})) \\
-(K_{a}+\Delta_{2}\cos(2\pi Q_{2}+\phi_{2})) & K_{a}+K_{b} & -(K_{b}+\Delta_{1}\cos(4\pi Q_{1}+\phi_{1}))  & \dots \\
\vdots & \vdots & \ddots & \vdots \\
-(K_{b}+\Delta_{1}\cos(2\pi Q_{1}+\phi_{1})) & \dots & -(K_{a}+\Delta_{2}\cos(2\pi Q_{2}+\phi_{2})) & K_{a}+K_{b} \\
\end{bmatrix}.
\end{equation}
}
\end{widetext}
The local topological marker is based on the eigenfunctions of the system and provides a local value for the topological invariant away from the boundaries. To construct a displacement matrix $U$, we organize all normalized eigenvectors $U_{j}$ with corresponding eigenfrequencies in ascending order. Thus, $U$ can be represented as $U=(U_{1},U_{2}, \ldots, U_{N})$. Consider $U_{-}=(U_{1},U_{2}, \ldots, U_{n+1})$ and $U_{+}=(U_{n+1},U_{n+2}, \ldots, U_{N})$, where the projectors below and above the band gap are denoted as $P_{-}=U_{-}U_{-}^{T}$ and $P_{+}=U_{+}U_{+}^{T}$. The flat band Hamiltonian is labeled as $Q=P_{+}-P_{-}$, and this can be further broken down into $Q=\tau_{A}Q\tau_{B}+\tau_{B}Q\tau_{A}$, through which the invariant can be expressed as:
\begin{equation}
    \nu(l)=\frac{1}{2}\sum_{a=A,B}((Q_{BA}[X,Q_{AB}])_{la,la}+(Q_{AB}[Q_{BA},X])_{la,la}).
\end{equation}
To calculate the winding number in a disordered system, we need to find the average of $\nu(l)$ over a small region at the central part of the chain for a single disorder instance. This process must be repeated for multiple disorder instances. In this context, $X$ represents the position operator, and l denotes the unit cell number.
The topological phase transition is associated with a diverging localization length at the Fermi level within the energy gap, $m\omega^{2}-m\omega_{0}^{2}=0$, given in Eq.~\eqref{eq:ANJU}
\begin{widetext}\label{eq:ANJU}
    \begin{equation}
    (K_{a}+\Delta_{2}\cos(2\pi Q_{2}j+\phi_{2}))u_{B}^{j}+(K_{b}+\Delta_{1}\cos(2\pi Q_{1}j+\phi_{1}))u_{B}^{j-1}=0,
\end{equation}
\end{widetext}

\begin{widetext}
\begin{equation}
    (K_{a}+\Delta_{2}\cos(2\pi Q_{2}j+\phi_{2}))u_{A}^{j}+(K_{b}+\Delta_{1}\cos(2\pi Q_{1}(j+1)+\phi_{1}))u_{A}^{j+1}=0.
\end{equation}
\end{widetext}

So simplifying the above expressions 
\begin{equation}
    u_{A}^{J+1}=-\frac{(K_{a}+\Delta_{2}\cos(2\pi Q_{2}j+\phi_{2}))u_{A}^{j}}{K_{b}+\Delta_{1}\cos(2\pi Q_{1}(j+1)+\phi_{1})}.
\end{equation}
\begin{equation}
     u_{A}^{N}=(-1)^{N-1}\prod_{j=1}^{j=N-1} \frac{(K_{a}+\Delta_{2}\cos(2\pi Q_{2}j+\phi_{2}))u_{A}^{1}}{K_{b}+\Delta_{1}\cos(2\pi Q_{1}(j+1)+\phi_{1})}.
\end{equation}
The Lyapunov exponent can be calculated as 
\begin{equation}
    \gamma=-\lim\limits_{N \to \infty}\frac{1}{N}\ln\left(\frac{(K_{a}+\Delta_{2}\cos(2\pi Q_{2}j+\phi_{2}))}{K_{b}+\Delta_{1}\cos(2\pi Q_{1}(j+1)+\phi_{1})}\right).
\end{equation}
\subsection{Localization length in commensurate case}

\begin{widetext}
    \begin{equation}
    \gamma_{comm}=-\lim\limits_{N \to \infty}\frac{1}{N}\ln\left(\frac{(\prod_{j=1}^{j=q_{2}}(K_{a}+\Delta_{2}\cos(2\pi\frac{p_{2}}{q_{2}}j+\phi_{2})))^\frac{N-1}{q_{2}}}{(\prod_{j=1}^{j=q_{1}}(K_{b}+\Delta_{1}\cos(2\pi\frac{p_{1}}{q_{1}}(j+1)+\phi_{1})))^\frac{N-1}{q_{1}}}\right),
\end{equation}
\end{widetext}
\begin{widetext}
    \begin{equation}
    \gamma_{comm}=-\lim\limits_{N \to \infty}\frac{1}{N}\ln\left(\left(\prod_{j=1}^{j=q_{2}}(K_{a}+\Delta_{2}\cos(2\pi\frac{p_{2}}{q_{2}}j+\phi_{2}))\right)^{a}\right)+\lim\limits_{N \to \infty}\frac{1}{N}\ln\left(\left(\prod_{j=1}^{j=q_{1}}(K_{b}+\Delta_{1}\cos(2\pi\frac{p_{1}}{q_{1}}(j+1)+\phi_{1}))\right)^{b}\right),
\end{equation}
\end{widetext}
where $b=\frac{N-1}{q_{1}}$, and $a=\frac{N-1}{q_{2}}$. Simplifying further,
\begin{widetext}
\begin{equation}
    \gamma_{comm}=-\frac{1}{q_{2}}\ln\left(\prod_{j=1}^{j=q_{2}}(K_{a}+\Delta_{2}\cos(2\pi\frac{p_{2}}{q_{2}}j+\phi_{2}))\right)+\frac{1}{q_{1}}\ln\left(\prod_{j=1}^{j=q_{1}}(K_{b}+\Delta_{1}\cos(2\pi\frac{p_{1}}{q_{1}}(j+1)+\phi_{1}))\right).
\end{equation}
\end{widetext}
The curve with $\gamma_{comm}=0$ is given as 
\begin{widetext}
    \begin{equation}
    \frac{q_{2}}{q_{1}}=\frac{\ln(\prod_{j=1}^{j=q_{2}}(K_{a}+\Delta_{2}\cos(2\pi\frac{p_{2}}{q_{2}}j+\phi_{2})))}{\ln(\prod_{j=1}^{j=q_{1}}(K_{b}+\Delta_{1}\cos(2\pi\frac{p_{1}}{q_{1}}(j+1)+\phi_{1})))}.
    \end{equation}
\end{widetext}

\subsection{Calculation of equations for boundary modes separating localized and delocalized modes}\label{subsec:2}
To calculate the mobility edges one has to include explicitly the stiffness constants of on-site springs as well
\begin{widetext}
    \begin{equation}
\begin{split}
    m\omega^{2}u_{A}^{j} &= (K_{a}+K_{b}+\Delta_{1}\cos(2\pi Q_{1}j+\phi_{1})+\Delta_{2}\cos(2\pi Q_{2}j+\phi_{2}))u_{A}^{j} \\
    &\quad - (K_{a}+\Delta_{2}\cos(2\pi Q_{2}j+\phi_{2}))u_{B}^{j} \\
    &\quad - (K_{b}+\Delta_{1}\cos(2\pi Q_{1}j+\phi_{1}))u_{B}^{j-1},
\end{split}
\end{equation}
\end{widetext}
\begin{widetext}
    \begin{equation}\label{eq:26}
\begin{split}
    m\omega^{2}u_{B}^{j} &= (K_{a}+K_{b}+\Delta_{1}\cos(2\pi Q_{1}(j+1)+\phi_{1})+\Delta_{2}\cos(2\pi Q_{2}j+\phi_{2}))u_{B}^{j} \\
    &\quad - (K_{a}+\Delta_{2}\cos(2\pi Q_{2}j+\phi_{2}))u_{A}^{j} \\
    &\quad - (K_{b}+\Delta_{1}\cos(2\pi Q_{1}(j+1)+\phi_{1}))u_{A}^{j+1}.
\end{split}
\end{equation}
\end{widetext}

Those equations can be simplified as
\begin{widetext}\label{eq:dft1}
    \begin{equation}
    u_{B}^{j}=\frac{(K_{a}+\Delta_{2}\cos(2\pi Q_{2}j+\phi_{2}))u_{A}^{j}+(K_{b}+\Delta_{1}\cos(2\pi Q_{1}(j+1)+\phi_{1}))u_{A}^{j+1}}{(K_{a}+K_{b}-m\omega^{2})+\Delta_{1}\cos(2\pi Q_{1}(j+1)+\phi_{1})+\Delta_{2}\cos(2\pi Q_{2}j+\phi_{2})},
\end{equation}
\end{widetext}

\begin{widetext}\label{eq:dft2}
\begin{equation}
    u_{B}^{j-1}=\frac{(K_{a}+\Delta_{2}\cos(2\pi Q_{2}j+\phi_{2}))u_{A}^{j-1}+(K_{b}+\Delta_{1}\cos(2\pi Q_{1}(j)+\phi_{1}))u_{A}^{j}}{(K_{a}+K_{b}-m\omega^{2})+\Delta_{1}\cos(2\pi Q_{1}(j)+\phi_{1})+\Delta_{2}\cos(2\pi Q_{2}(j-1)+\phi_{2})}.
\end{equation}
\end{widetext}

Substituting $X=m\omega^{2}-(K_{a}+K_{b})$, and
\begin{equation}
    h_{j}^{\Delta_{1}}=\Delta_{1}\cos(2\pi Q_{1}j+\phi_{1}),
\end{equation}
\begin{equation}
    h_{j}^{\Delta_{2}}=\Delta_{2}\cos(2\pi Q_{2}j+\phi_{2}),
\end{equation} into Eq.~\eqref{eq:dft1} and Eq.~\eqref{eq:dft2}
\begin{equation}
 u_{B}^{j}=\frac{(K_{a}+h_{j}^{\Delta_{2}})u_{A}^{j}+(K_{b}+h_{j+1}^{\Delta_{1}})u_{A}^{j+1}}{-X+h_{j+1}^{\Delta_{1}}+h_{j}^{\Delta_{2}}},
\end{equation}
\begin{equation}
 u_{B}^{j-1}=\frac{(K_{a}+h_{j-1}^{\Delta_{2}})u_{A}^{j-1}+(K_{b}+h_{j}^{\Delta_{1}})u_{A}^{j}}{-X+h_{j}^{\Delta_{1}}+h_{j-1}^{\Delta_{2}}}.
\end{equation}
Plugging these equations into Eq.~\eqref{eq:26}, 
\begin{widetext}
\begin{equation}
\begin{split}
    u_{A}^{j+1} &= \frac{(X-h_{j}^{\Delta_{1}}-h_{j}^{\Delta_{2}})^{2}(X-h_{j+1}^{\Delta_{1}}-h_{j}^{\Delta_{2}})-(K_{a}+h_{j}^{\Delta_{2}})^{2}(X-h_{j}^{\Delta_{1}}-h_{j}^{\Delta_{2}})}{(K_{a}+h_{j}^{\Delta_{2}})(K_{b}+h_{j+1}^{\Delta_{1}})(X-h_{j}^{\Delta_{1}}-h_{j}^{\Delta_{2}})}u_{A}^{j} \\
    & \quad -\frac{(K_{a}+h_{j-1}^{\Delta_{2}})(K_{b}+h_{j}^{\Delta_{1}})(X-h_{j+1}^{\Delta_{1}}-h_{j}^{\Delta_{2}})}{(K_{a}+h_{j}^{\Delta_{2}})(K_{b}+h_{j+1}^{\Delta_{1}})(X-h_{j}^{\Delta_{1}}-h_{j-1}^{\Delta_{2}})}u_{A}^{j-1}.
\end{split}
\end{equation}
\end{widetext}

To calculate the localization length, the transfer matrix is calculated as 

\begin{equation}
    T^{j}(\omega,\phi_{1},\phi_{2})=
\begin{bmatrix}
    a & d\\
    1 & 0
\end{bmatrix},
\end{equation}

where 
\begin{widetext}
    \begin{equation}
    a=\frac{(X-h_{j}^{\Delta_{1}}-h_{j}^{\Delta_{2}})^{2}(X-h_{j+1}^{\Delta_{1}}-h_{j}^{\Delta_{2}})-(K_{a}+h_{j}^{\Delta_{2}})^{2}(X-h_{j}^{\Delta_{1}}-h_{j}^{\Delta_{2}})-(K_{b}+h_{j}^{\Delta_{1}})^{2}(X-h_{j+1}^{\Delta_{1}}-h_{j}^{\Delta_{2}})}{(K_{a}+h_{j}^{\Delta_{2}})(K_{b}+h_{j+1}^{\Delta_{1}})(X-h_{j}^{\Delta_{1}}-h_{j}^{\Delta_{2}})},
\end{equation}
\end{widetext}
\begin{widetext}
    \begin{equation}
        d=\frac{(K_{a}+h_{j-1}^{\Delta_{2}})(K_{b}+h_{j}^{\Delta_{1}})(X-h_{j+1}^{\Delta_{1}}-h_{j}^{\Delta_{2}})}{(K_{a}+h_{j}^{\Delta_{2}})(K_{b}+h_{j+1}^{\Delta_{1}})(X-h_{j}^{\Delta_{1}}-h_{j-1}^{\Delta_{2}})}.
    \end{equation}
\end{widetext}

This transfer matrix can be decomposed as 
\begin{equation}
    T^{j}(\omega,\phi_{1},\phi_{2})=U^{j}(\omega,\phi_{1},\phi_{2})V^{j}(\omega,\phi_{1},\phi_{2})
\end{equation}
Where 
\begin{equation}
    U^{j}(\omega,\phi_{1},\phi_{2})=\frac{(K_{a}+h_{j}^{\Delta_{2}})^{2}}{(K_{a}+h_{j}^{\Delta_{2}})(K_{b}+h_{j+1}^{\Delta_{1}})},
\end{equation} and

\begin{widetext}
    \begin{equation}\label{eq:B}
    V^{j}(\omega,\phi_{1},\phi_{2}) =
    \begin{bmatrix}
        G
        &
        \frac{(K_{a}+h_{j-1}^{\Delta_{2}})(K_{b}+h_{j}^{\Delta_{1}})(X-h_{j+1}^{\Delta_{1}}-h_{j}^{\Delta_{2}})}{(K_{a}+h_{j}^{\Delta_{2}})^{2}(X-h_{j}^{\Delta_{1}}-h_{j-1}^{\Delta_{2}})} \\
        \frac{(K_{a}+h_{j}^{\Delta_{2}})(K_{b}+h_{j+1}^{\Delta_{1}})}{(K_{a}+h_{j}^{\Delta_{2}})^{2}} & 0
    \end{bmatrix},
\end{equation}
\end{widetext}
where 
\begin{widetext}
    \begin{equation}
        G=\frac{(X-h_{j}^{\Delta_{1}}-h_{j}^{\Delta_{2}})^{2}(X-h_{j+1}^{\Delta_{1}}-h_{j}^{\Delta_{2}})}{(K_{a}+h_{j}^{\Delta_{2}})^{2}(X-h_{j}^{\Delta_{1}}-h_{j}^{\Delta_{2}})} -\frac{(K_{a}+h_{j}^{\Delta_{2}})^{2}(X-h_{j}^{\Delta_{1}}-h_{j}^{\Delta_{2}})-(K_{b}+h_{j}^{\Delta_{1}})^{2}(X-h_{j+1}^{\Delta_{1}}-h_{j}^{\Delta_{2}})}{(K_{a}+h_{j}^{\Delta_{2}})^{2}(X-h_{j}^{\Delta_{1}}-h_{j}^{\Delta_{2}})}.
    \end{equation}
\end{widetext}
So the Lyapunov exponent can be calculated in separate form as 
\begin{equation}
    \gamma(\omega)=\gamma_{U}(\omega)+\gamma_{V}(\omega)
\end{equation}
Where 
\begin{equation}\label{eq:Ly2}
    \gamma_{U}(\omega)=\lim\limits_{N \to \infty}\frac{1}{N}\ln\lVert \prod_{j=1}^{j=N}U_{j}(\omega,\phi_{1},\phi_{2}) \rVert,
\end{equation}
\begin{equation}\label{eq:Ly1}
    \gamma_{V}(\omega)=\lim\limits_{N \to \infty}\frac{1}{N}\ln\lVert \prod_{j=1}^{j=N}V_{j}(\omega,\phi_{1},\phi_{2}) \rVert.
\end{equation}
Taking only the term with $\phi_{1}$ one can apply ergodic theory to solve such equation 
\begin{equation}
    \gamma_{U}(\omega)=\frac{1}{2\pi}\int_{0}^{2\pi}\ln\left|\frac{K_{a}^{2}}{K_{a}(K_{b}+\Delta_{1}\cos(\eta))}\right|d\eta .
\end{equation}
The solution for $K_{b}>\Delta_{1}$ is
\begin{equation}
    \gamma_{U}(\omega)=-\ln\left(\frac{K_{b}+\sqrt{K_{b}^{2}-\Delta_{1}^{2}}}{2K_{a}}\right),
\end{equation}
and for $K_{b}<\Delta_{1}$
\begin{equation}
    \gamma_{U}(\omega)=-\ln\left(\frac{\Delta_{1}}{2K_{a}}\right).
\end{equation}
For calculating $\gamma_{V}(\omega)$ using $V^{j}(\omega,\phi)$, we introduced imaginary term in the phase of the cosine term, i.e $\phi=\phi+ic$ and take that imaginary part $c$ very large, in that limit the term becomes
\begin{equation}
    \frac{X-f_{j+1}}{X-f_{j}}=e^{-2\pi i\beta},
\end{equation}
and the expression for $h_{j}^{\Delta_{1}}=\frac{\Delta_{1}}{2}e^{-(2\pi i \beta j+i\phi)}e^{c}$. Putting all these in Eq.~\eqref{eq:B} we have the final expression for $V^{j}$  as 
\begin{equation}
    V^{j}=e^{-(2\pi i\beta(j+1)+i\phi)+c}Z(\omega,\Delta_{1}),
\end{equation}
where the $2\times2$ matrix $Z$ is 
\begin{equation}
    Z(\omega,\Delta_{1})=\begin{pmatrix}
  \frac{-\Delta_{1}(m\omega^{2}-K_{a})}{K_{a}^{2}} & \frac{-\Delta_{1}}{2K_{a}} \\
  \frac{\Delta_{1}}{2K_{a}} & 0
\end{pmatrix}.
\end{equation}
From Eq.~\eqref{eq:Ly1} the lyapunov exponent will be as follows
\begin{equation}
    \gamma_{V}(\omega,c)=c+\ln\left|\frac{\Delta_{1}(K_{a}-m\omega^{2}+\sqrt{m\omega^{2}(m\omega^{2}-2K_{a})}}{2K_{a}^{2}}\right|.
\end{equation}
Following \cite{10.1007/s11511-015-0128-7}, one can write
\begin{equation}
    \gamma_{V}(\omega)=\ln\left|\frac{\Delta_{1}(K_{a}-m\omega^{2}+\sqrt{m\omega^{2}(m\omega^{2}-2K_{a})}}{2K_{a}^{2}}\right|.
\end{equation}
\begin{figure*}[htbp]
    \centering
    \includegraphics[keepaspectratio, width=0.45\textwidth]{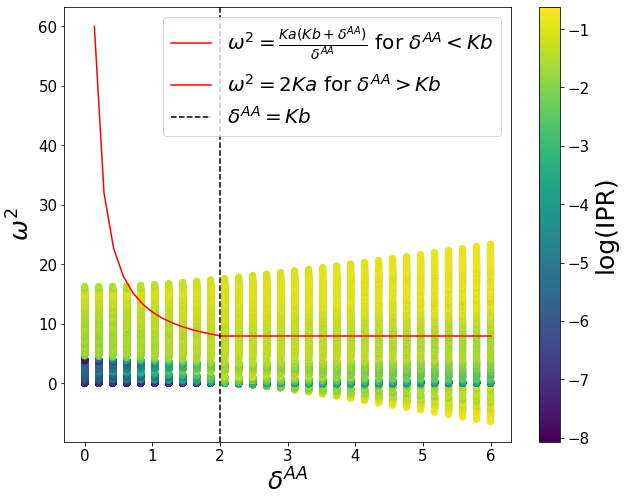}
    \includegraphics[keepaspectratio, width=0.45\textwidth]{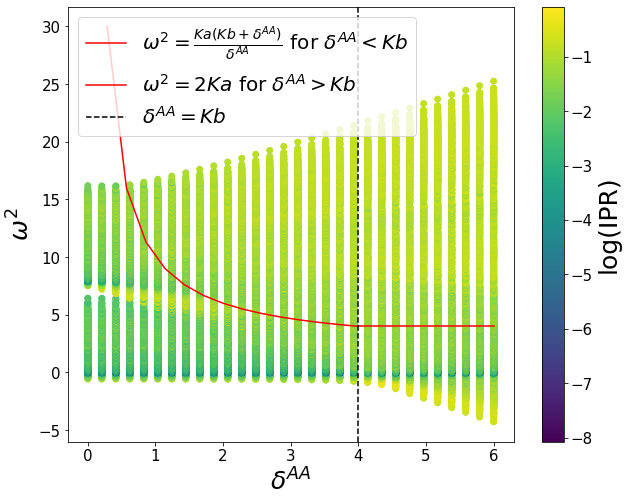}
     \put(-380,100){\textbf{(a) }}
     \put(-150,100){\textbf{(b)}}
     \\
     \includegraphics[keepaspectratio, width=0.45\textwidth]{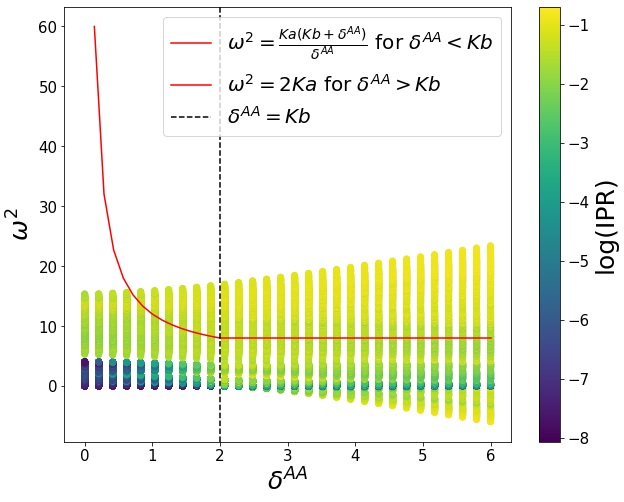}
    \includegraphics[keepaspectratio, width=0.45\textwidth]{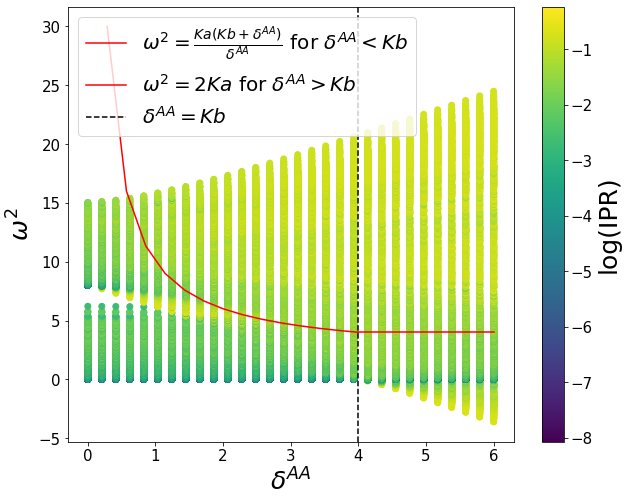}
     \put(-380,100){\textbf{(c) }}
     \put(-150,100){\textbf{(d)}}
     \\
     \caption{ Plot of IPR as a function of eigenvalues of the non-chiral model for $N=1597$ unit cells, $\phi=0$, and $\delta_{d}>2$. (a): $K_{a}>K_{b}$ for $\delta_{d}=2.5$, (b): $K_{a}<K_{b}$ for $\delta_{d}=2.5$, (c): $k_{a}>K_{b}$ for $\delta_{d}=2$, (d): $K_{a}<K_{b}$ for $\delta_{d}=2$.  The figures above show that as $\delta_{d}$ increases, the presence of localized, extended, and critical states within well-defined boundaries diminishes, with the states becoming progressively localized and somewhat in a critical phase, leading to a disappearance of mobility edges due to discrete chiral disorder. In cases (a) and (c), critical states appear below the critical strength of quasi-periodic modulation $\delta^{AA}$ with a new analytic expression for the mobility edge, which cannot be accounted for, using Ref.~\cite{10.1007/s11511-015-0128-7}. }
    \label{fig:extra}
\end{figure*}

Hence the total lyapunov exponent is for $K_{b}>\Delta_{1}$
\begin{equation}
    \gamma(\omega)=\ln\left(\left|\frac{\Delta_{1}(K_{a}-m\omega^{2}+\sqrt{m\omega^{2}(m\omega^{2}-2K_{a})})}{K_{a}(K_{b}+\sqrt{K_{b}^{2}-\Delta_{1}^{2}})}\right|\right),
\end{equation}
for $K_{b}<\Delta_{1}$ 
\begin{equation}
    \gamma(\omega)=\ln\left(\left|\frac{\Delta_{1}(K_{a}-m\omega^{2}+\sqrt{m\omega^{2}(m\omega^{2}-2K_{a})})}{K_{a}}\right|\right).
\end{equation}
The equation for mobility edges for the  case $K_{b}<\Delta_{1}$ will be 
\begin{equation}
    m\omega^{2}=2K_{a},
\end{equation} and for $K_{b}>\Delta_{1}$ we have 
\begin{equation}\label{eq:f}
    m\omega^{2}=\frac{K_{a}(K_{b}+\Delta_{1})}{\Delta_{1}}.
\end{equation}
The above equations are plotted in Fig.~\ref{fig:Nonchiral} and Fig.~\ref{fig:extra}.

\begin{figure*}[htbp]
    \centering
    \begin{subfigure}[b]{0.4\linewidth}
        \centering
        \includegraphics[width=\linewidth]{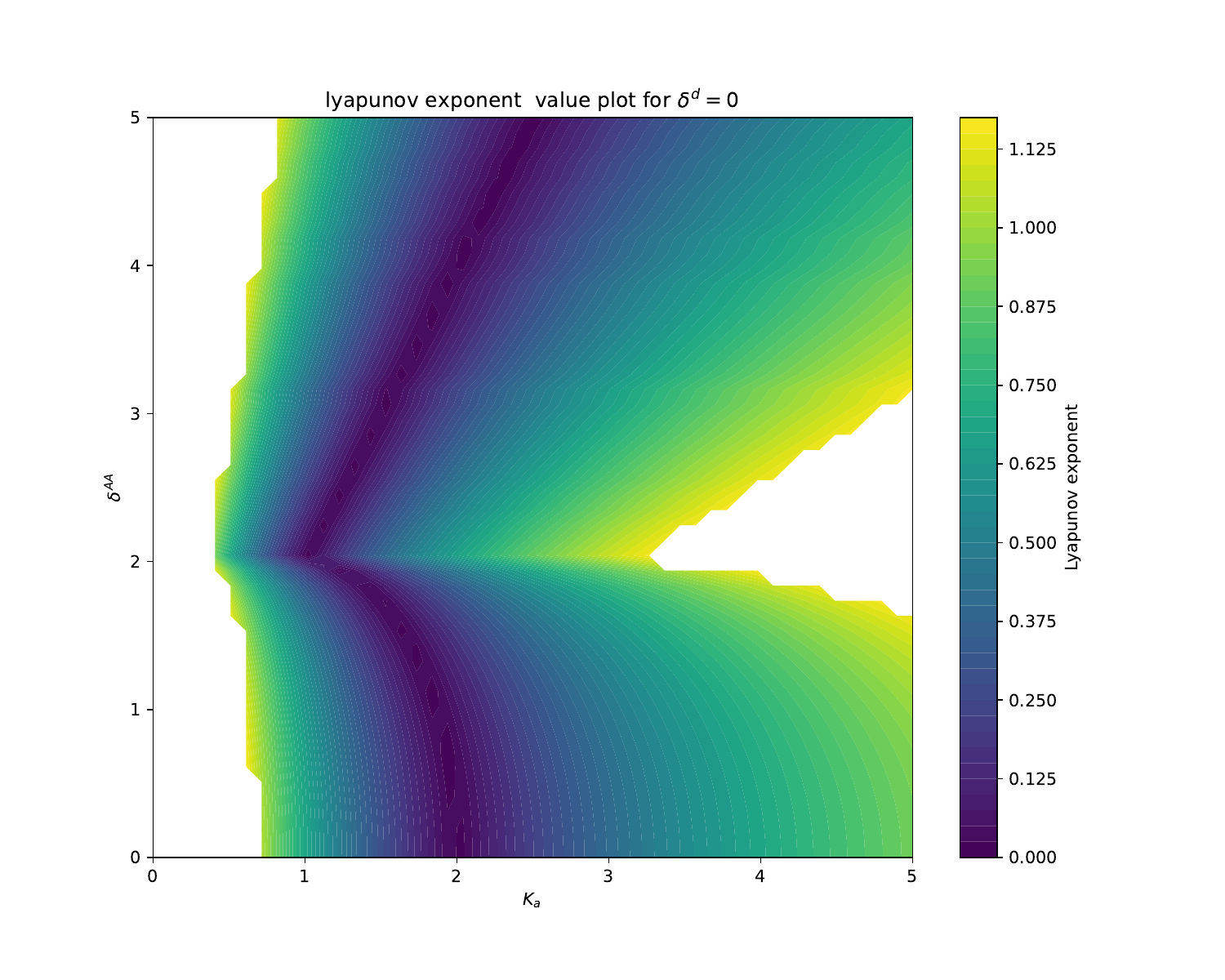}
        \caption{$\delta_{d}=0$ in incommensurate modulation}
        \label{fig:subfig1A}
    \end{subfigure}
    \begin{subfigure}[b]{0.4\linewidth}
        \centering
        \includegraphics[width=\linewidth]{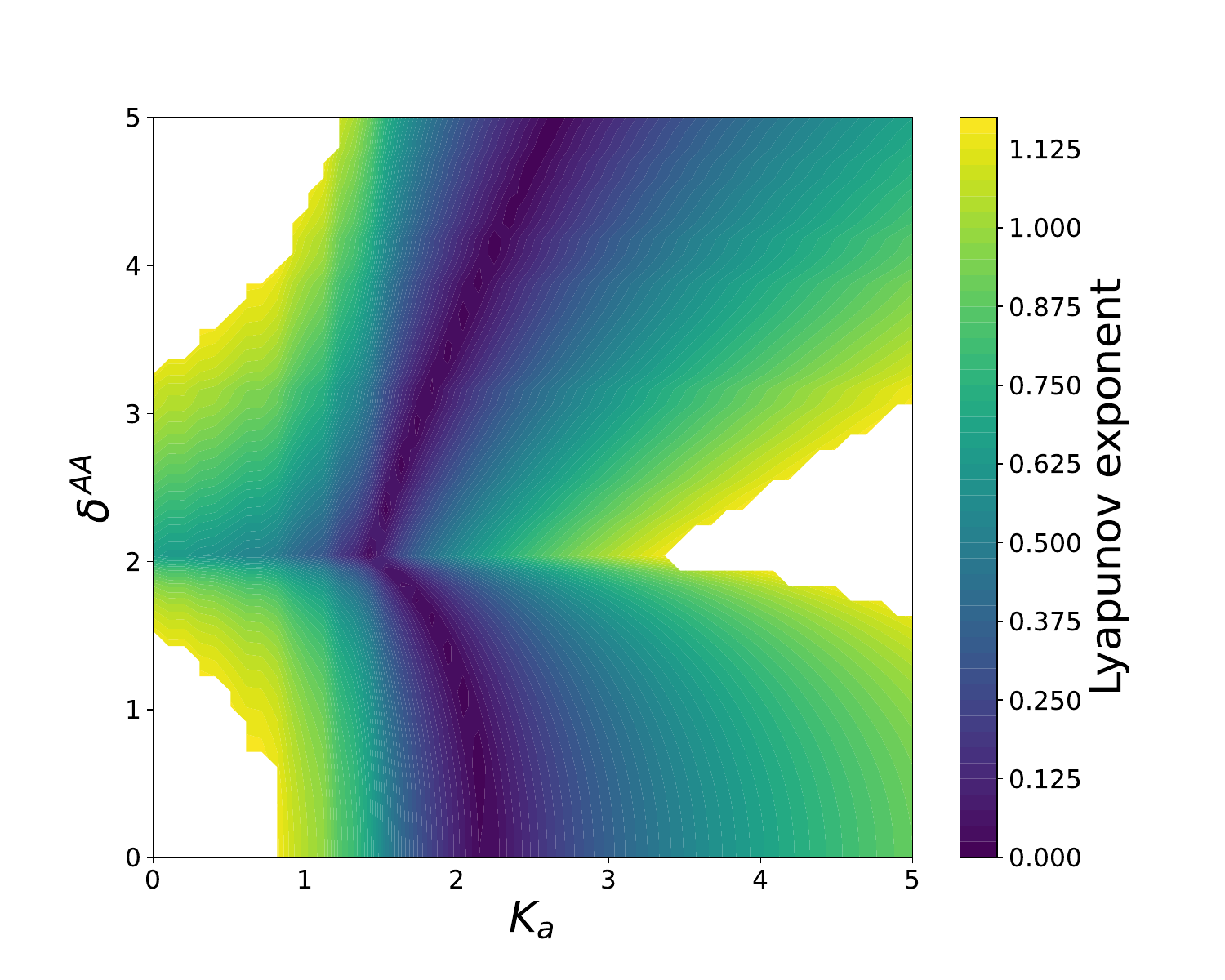}
        \caption{$\delta_{d}=1.5$ in incommensurate modulation}
        \label{fig:subfig2A}
    \end{subfigure}
    \begin{subfigure}[b]{0.4\linewidth}
        \centering
        \includegraphics[width=\linewidth]{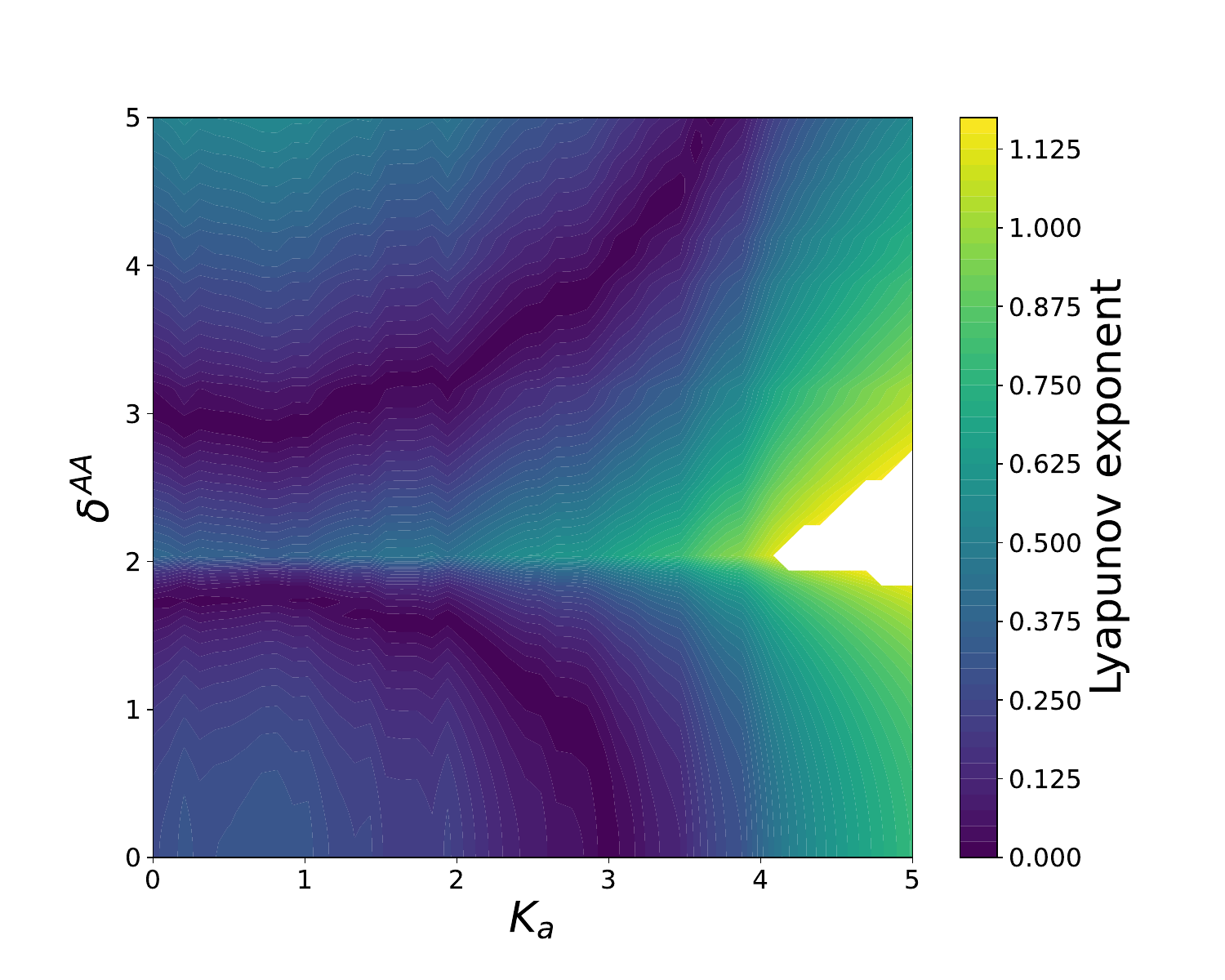}
        \caption{$\delta_{d}=4$ in incommensurate modulation}
        \label{fig:subfig3A}
    \end{subfigure}
    \begin{subfigure}[b]{0.4\linewidth}
        \centering
        \includegraphics[width=\linewidth]{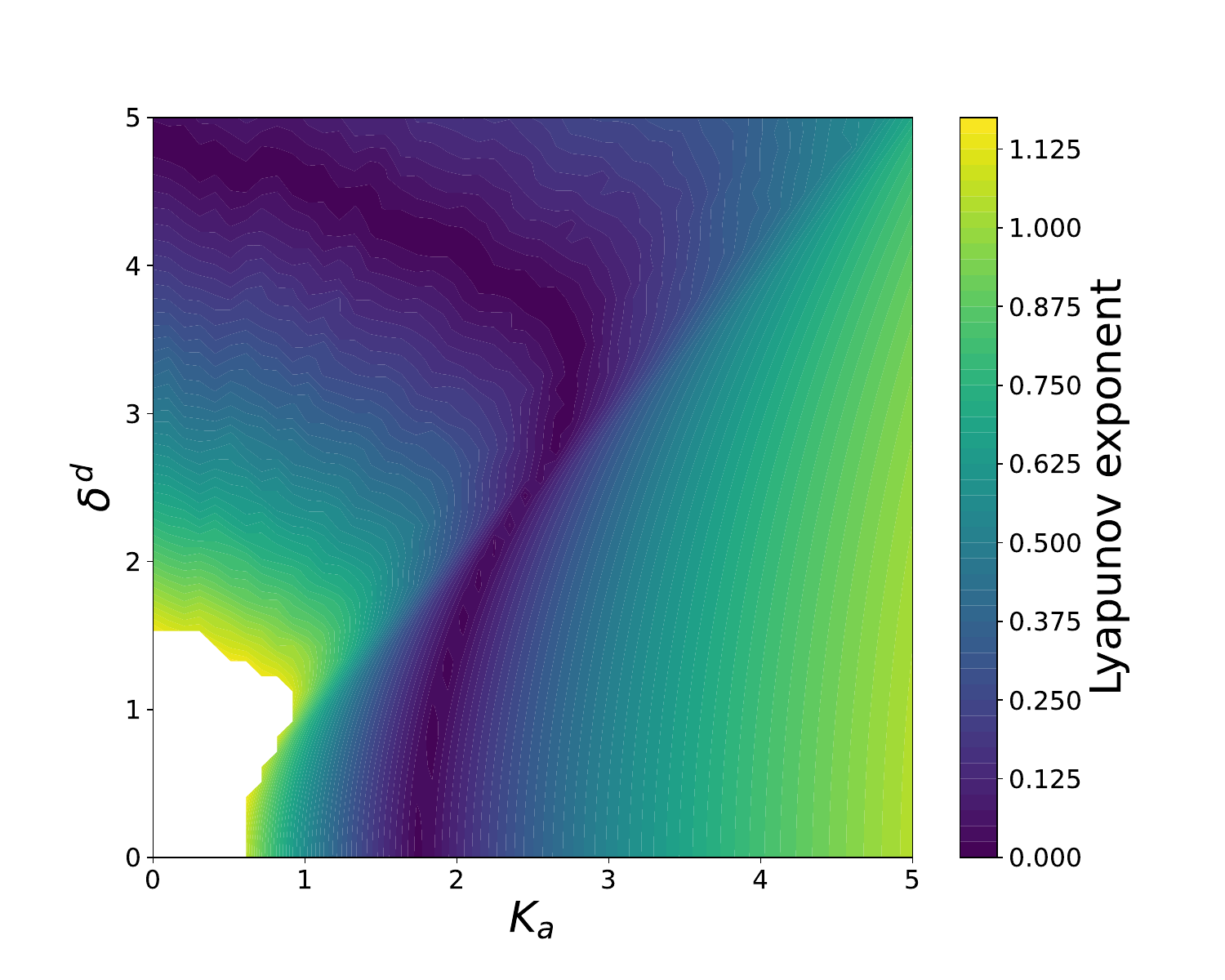}
        \caption{$\delta^{AA}=1.3$ in incommensurate modulation}
        \label{fig:subfig4A}
    \end{subfigure}
    \begin{subfigure}[b]{0.4\linewidth}
        \centering
        \includegraphics[width=\linewidth]{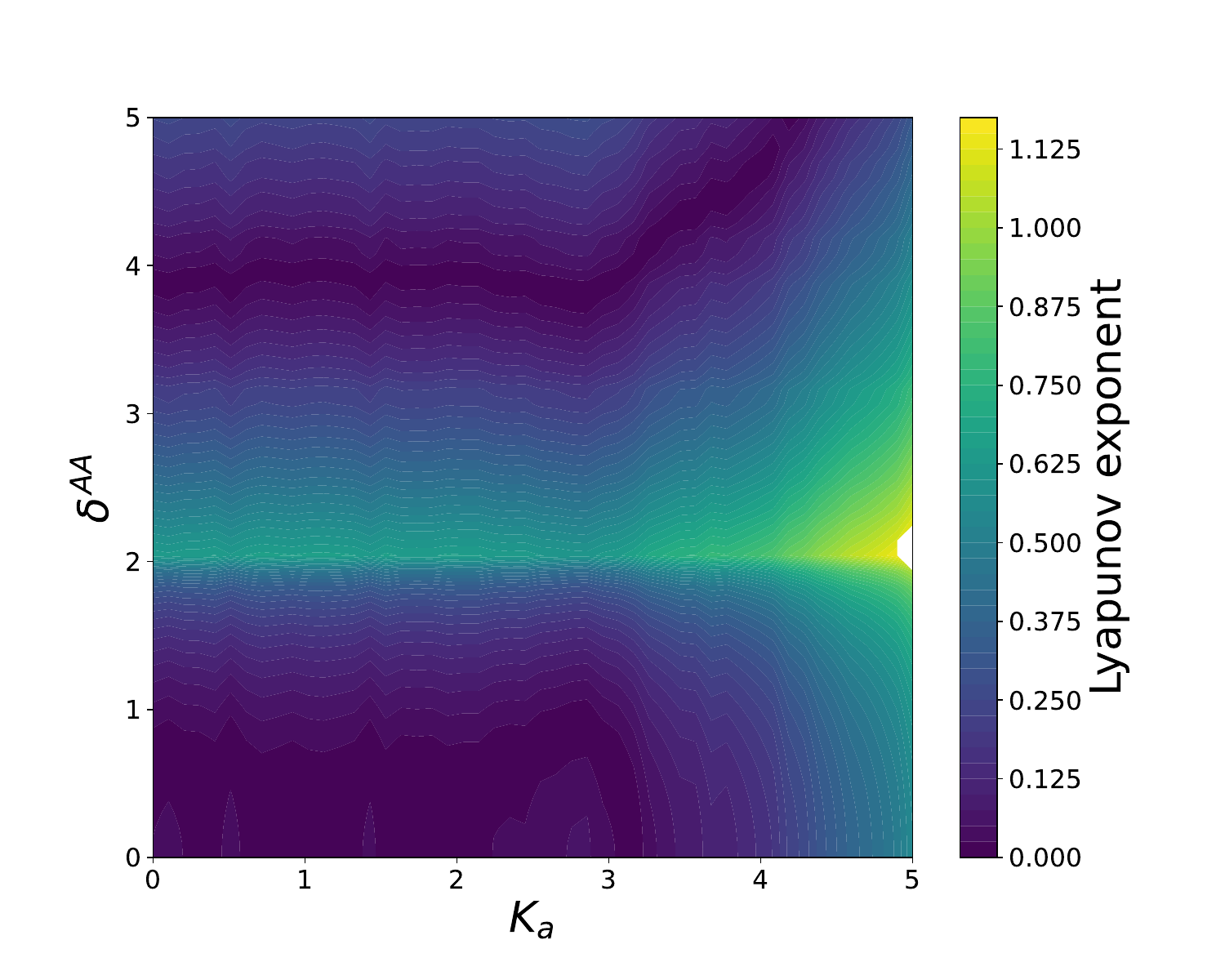}
        \caption{$\delta_{d}=5$ in incommensurate modulation}
        \label{fig:subfig5A}
    \end{subfigure}
    \begin{subfigure}[b]{0.4\linewidth}
        \centering
        \includegraphics[width=\linewidth]{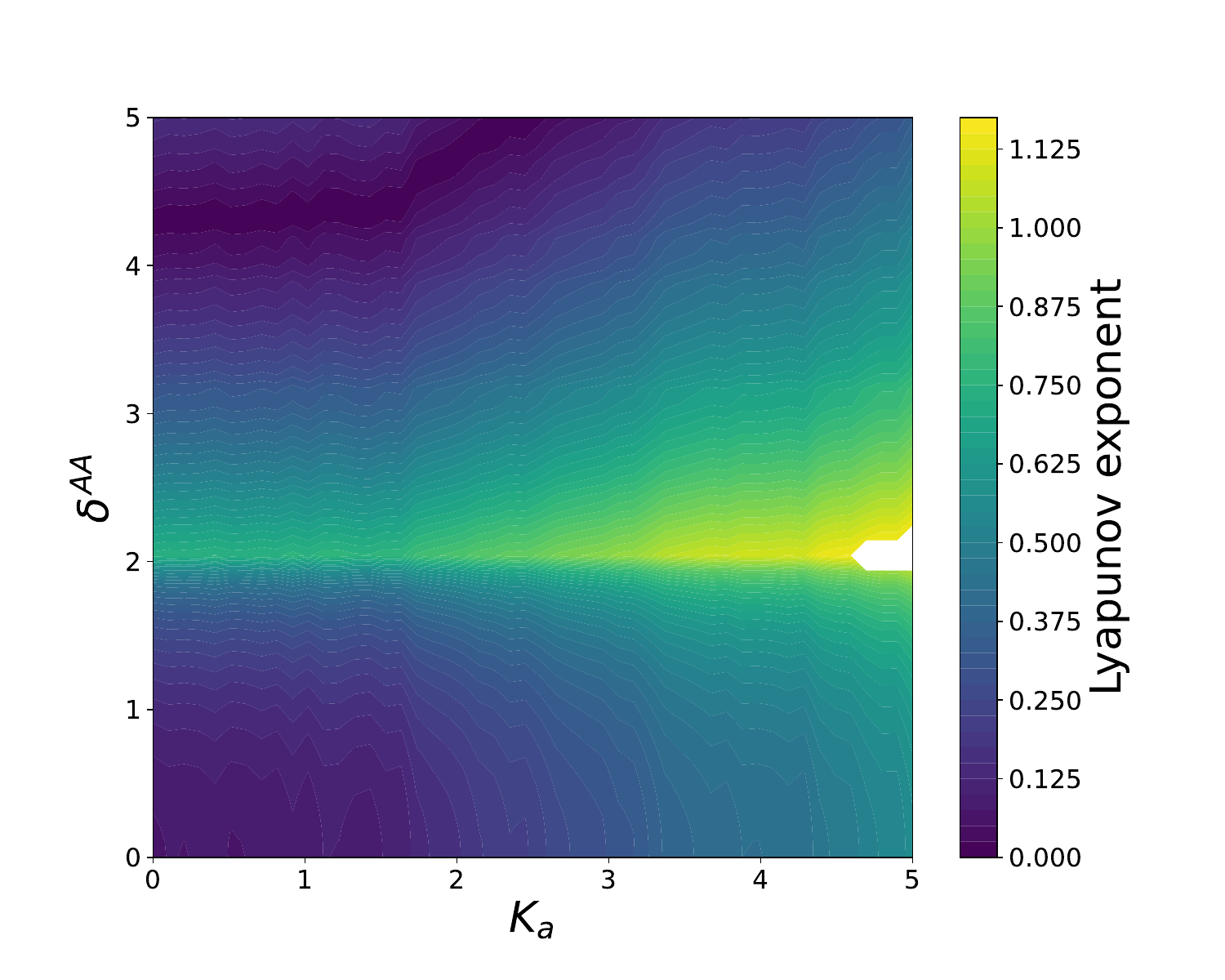}
        \caption{$\delta_{d}=6$ in incommensurate modulation}
        \label{fig:subfig6A}
    \end{subfigure}
    \caption{Plot of Lyapunov exponent in the phase space of parameters $K_{a}$ versus $\delta^{AA}$ in figures (a), (b), (c), (e), (f), and $K_{a}$ vs $\delta_{d}$ in figure (d). The dark blue region indicates diverging localization length because the Lyapunov exponent tends to $0$ along those curves, which represent the topological phase transition boundary. Figures (c), (e), and (f) are particularly interesting as they depict a re-entrant topological phase transition, starting from non-trivial phase, akin to what is shown in Fig.~\ref{fig:vector_plot97}. Fig.~\ref{fig:subfig4A} shows a unique re-entrant phase transition from a trivial phase to a non-trivial phase with distinct localization properties suggestive of a topological Anderson insulator. When considering negative values of $K_{a}$ and $\delta_{d}$, a rich two-way phase transition emerges, but in this context, focusing on only positive values of spring stiffness, the region of the topological Anderson insulator exists between $\delta_{d}>1$ and $\delta_{d}<4$. }
    \label{fig:mainC}
\end{figure*}

\begin{figure*}[htbp]
    \centering

    \begin{subfigure}{0.45\textwidth}
        \centering
        \includegraphics[width=\linewidth]{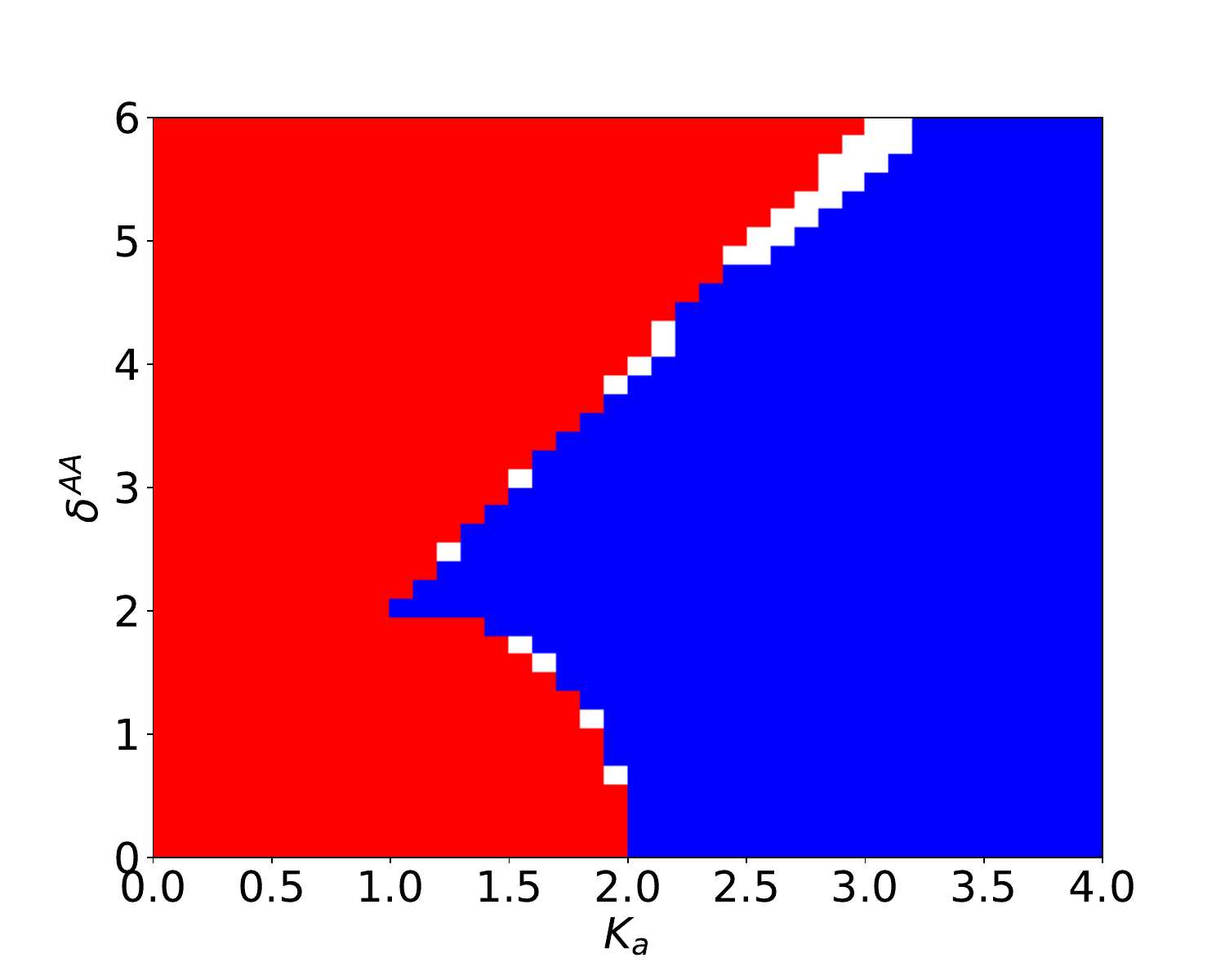}
        \caption{$\delta_{d}=0$, $\phi=0$}
        \label{fig:vector_plot90}
    \end{subfigure}
    \hspace{0.05\textwidth} 
    \begin{subfigure}{0.45\textwidth}
        \centering
        \includegraphics[width=\linewidth]{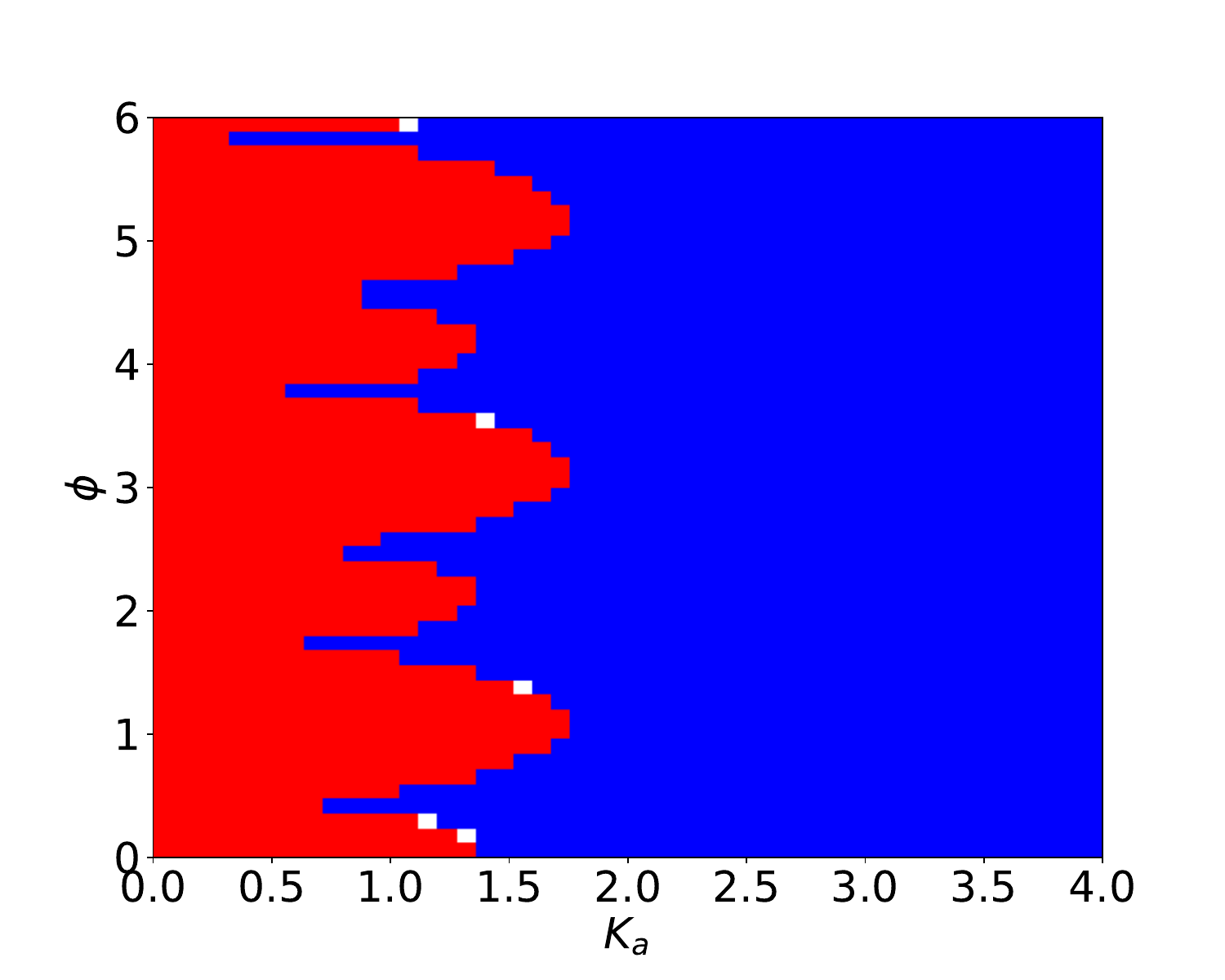}
        \caption{ $\delta^{AA}=\frac{1}{3}$,$\delta_{d}=0$}
        \label{fig:vector_plot91}
    \end{subfigure}

    \vspace{0.05\textwidth} 

    \begin{subfigure}{0.45\textwidth}
        \centering
        \includegraphics[width=\linewidth]{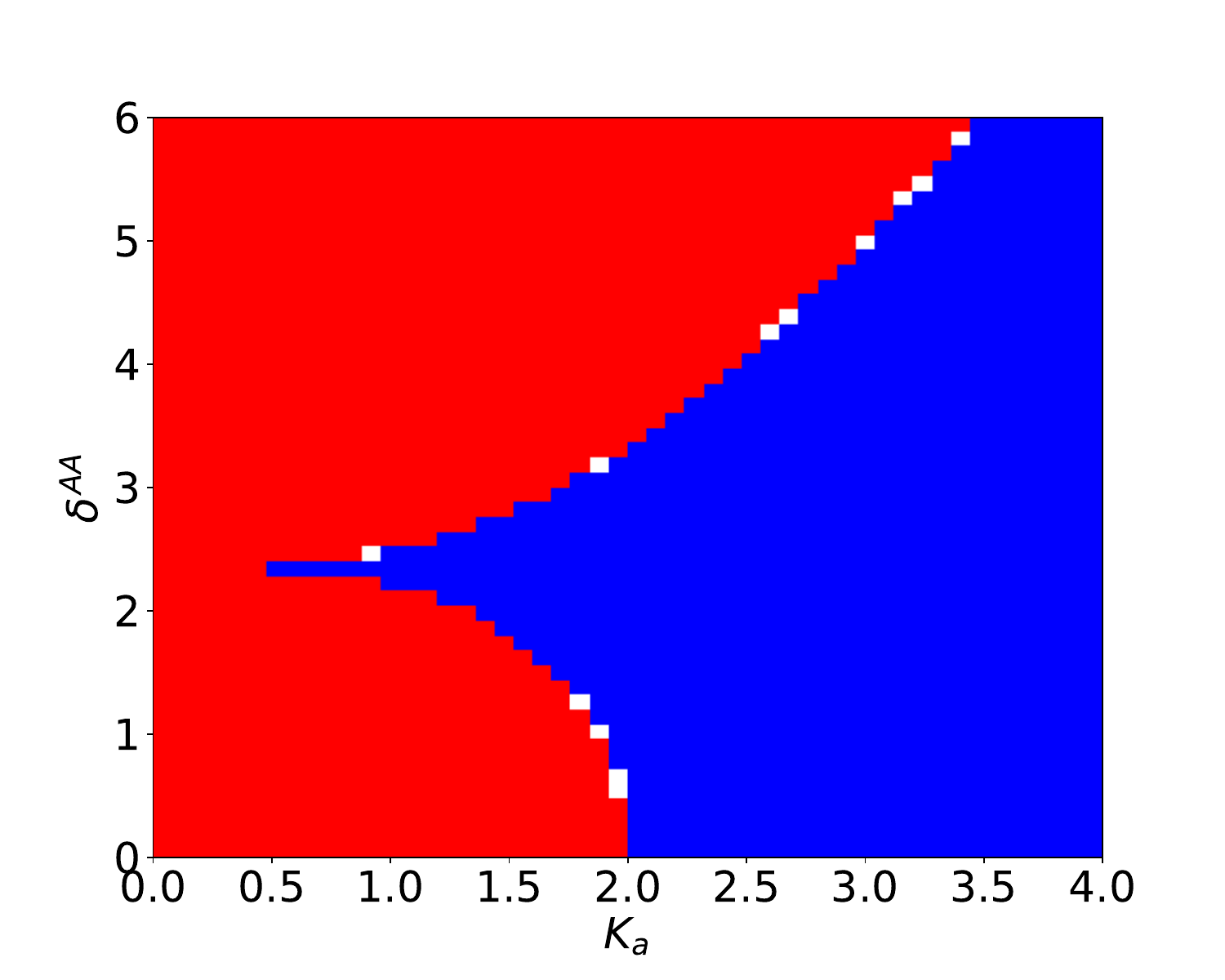}
        \caption{ $\delta_{d}=0$,$\phi=0.5$}
        \label{fig:vector_plot92}
    \end{subfigure}
    \hspace{0.05\textwidth} 
    \begin{subfigure}{0.45\textwidth}
        \centering
        \includegraphics[width=\linewidth]{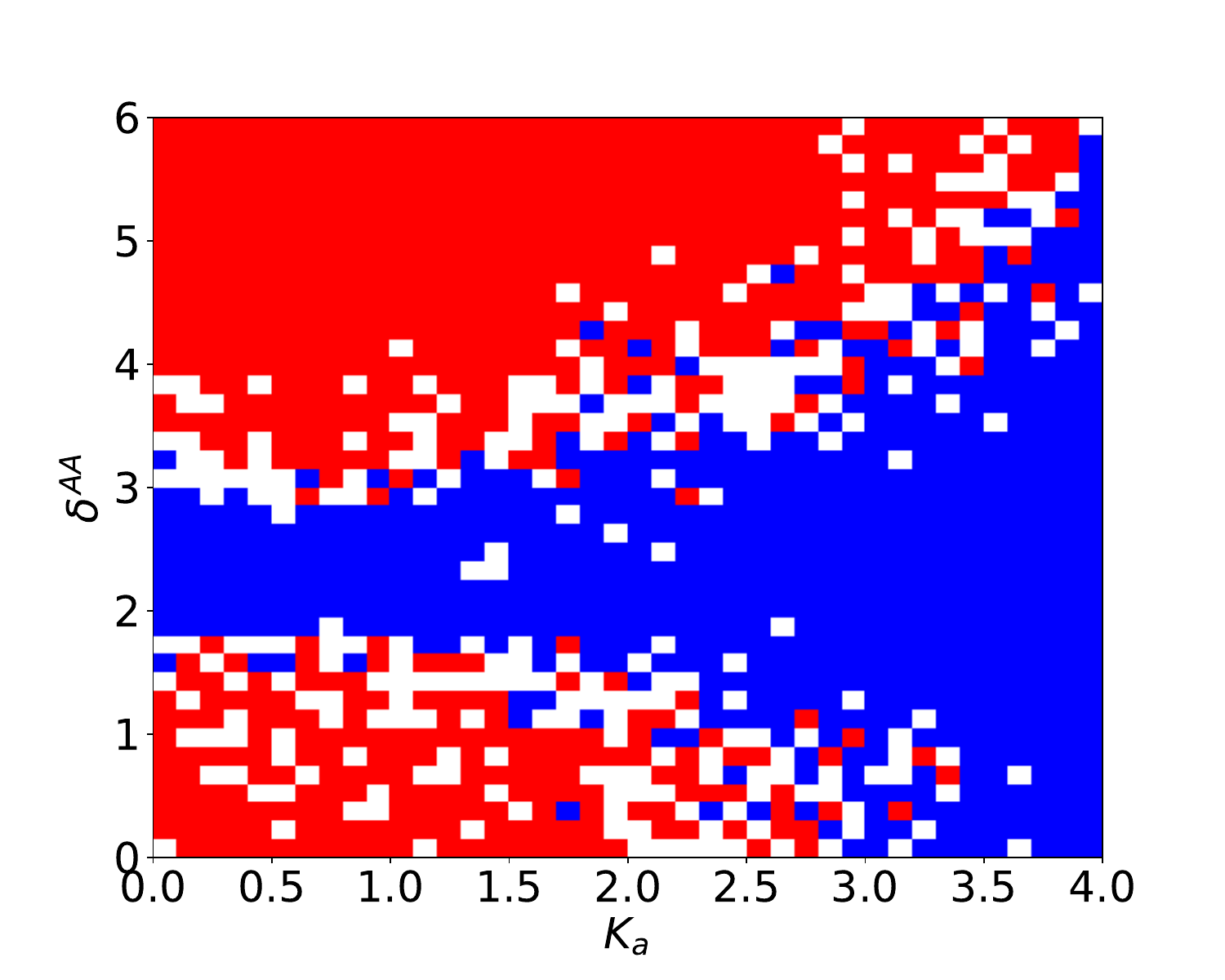}
        \caption{$\delta_{d}=4$, $\phi=0$}
        \label{fig:vector_plot97}
    \end{subfigure}

    \caption{Average LTM calculated for a chain of spring masses. Plots in (a) and (d) depict the incommensurate case, while (b) and (c) represent the commensurate scenario. The red area signifies a topological invariant of $1$, indicating a non-trivial topological phase, whereas the blue area shows an invariant of $0$, representing a trivial topological phase. The white strip separates the trivial and non-trivial phases, where the localization length diverges. The figures shows a bidirectional topological phase transition, with the divergence of the localization length. The interplay between chiral disorder and quasi-periodic modulation leads to a re-entrant behaviour of phase transition starting from a non-trivial phase. This transition culminates in a complete trivial phase within a specific range of $\delta^{AA}$ regardless of the parameter $K_{a}$, as illustrated in Fig.~\ref{fig:vector_plot97}. Furthermore, the transition from the trivial phase gives rise to a non-trivial topologically insulating Anderson phase, whose boundary is continually influenced by the strength of the quasi-periodic potential, as shown in Fig.~\ref{fig:subfig4A}. }
    \label{fig:combined_vector_plot80}
\end{figure*}

\section*{Acknowledgment}
The author acknowledge support of the Department of
Atomic Energy, Government of India, under
Project Identification No. RTI4007. The author also thank Dr. Kabir Ramola for reviewing the article.
\end{document}